\documentclass[aps,prd, twocolumn,superscriptaddress,floatfix,preprintnumbers,nofootinbib]{revtex4-2}
\usepackage[utf8]{inputenc}
\usepackage{xspace}
\usepackage{amsmath}
\usepackage{amssymb}
\usepackage{bm}
\usepackage{xcolor}
\usepackage{feynmf}
\usepackage{slashed}
\usepackage{braket}
\usepackage{url}
\usepackage{natbib}
\usepackage{graphicx}
\usepackage{subfigure}
\usepackage{mathrsfs}  
\usepackage{cancel}
\usepackage[normalem]{ulem}
\usepackage[pdftex,bookmarks,linktocpage,pdfpagelabels,plainpages=false,hyperfigures,linkcolor=blue,citecolor=blue]{hyperref}
\hypersetup{colorlinks=true}

\begin{document}

\title{Hints of Natural Supersymmetry in Flavor Anomalies?}

\author{P.~S.~Bhupal Dev}
\email{bdev@wustl.edu}
\affiliation{Department of Physics and McDonnell Center for the Space Sciences, Washington University, St. Louis, MO 63130, USA}
\author{Amarjit Soni}
\email{adlersoni@gmail.com}
\affiliation{Physics Department, Brookhaven National Laboratory, Upton, NY 11973, USA}
\author{Fang Xu}
\email{xufang@wustl.edu}
\affiliation{Department of Physics and McDonnell Center for the Space Sciences, Washington University, St. Louis, MO 63130, USA}

\begin{abstract}
The recent results from the Fermilab muon $g-2$ experiment, as well as the persisting hints of lepton flavor universality violation in $B$-meson decays, present a very strong case for  flavor-nonuniversal new physics beyond the Standard Model. We assert that a minimal $R$-parity violating supersymmetric scenario with relatively light third-generation sfermions (dubbed as `RPV3') provides a natural, well-motivated framework for the simultaneous explanation of all flavor anomalies, while being consistent with a multitude of low-energy flavor constraints, as well as with limits from high-energy collider searches. We further propose complementary tests and distinct signatures of this scenario in the high-$p_T$ searches at current and future colliders. Specifically, we find that an sbottom in the mass range of 2--12 TeV accounts for $R_{D^{(*)}}$ and $R_{K^{(*)}}$ flavor anomalies and it only plays a minor role in the $(g-2)_\mu$ anomaly, whereas a sneutrino  with mass between 0.7--1 TeV is the dominant player for $(g-2)_\mu$. In this context, we propose specific collider signatures of sbottom via its decays to $ \overline{ t}(t) \mu^+ \mu^-$, and of sneutrino pairs with their decays leading to a highly distinctive and spectacular four-muon final state, which can be used to completely probe the RPV3 parameter space of interest. 
\end{abstract}

\maketitle

%%%%%%%%%%%%%%%%%%%%%%%%%%%
\section{Introduction}\label{sec:intro}
This is an era of {\it anomalies}, as a growing list of experimental results, ranging from flavor physics to neutrinos to Dark Matter, show deviations from the Standard Model (SM) expectations at a few $\sigma$ level~\cite{Fischer:2021sqw}. While most of them might be an artifact (or a combination) of statistical fluctuations, systematic effects, theory/background uncertainties, experimental errors or other unknown issues that need further scrutiny, the intriguing possibility that some anomalies might represent genuine new physics signals makes it worthwhile studying all possible aspects, including beyond the SM (BSM) interpretations. 

Of the existing statistically significant ($\gtrsim 3\sigma$) anomalies, particularly striking are the persistent hints of Lepton Flavor Universality Violation (LFUV) in semileptonic $B$-meson decays, as reported by three different experiments with completely independent datasets, namely,  BaBar~\cite{Lees:2012xj}, Belle~\cite{Huschle:2015rga,Hirose:2016wfn,Abdesselam:2019dgh, Abdesselam:2019wac,Abdesselam:2019lab} and LHCb~\cite{Aaij:2015yra,Aaij:2017uff, Aaij:2017vbb, Aaij:2019wad, Aaij:2021vac}, with a combined significance of $4.5\sigma$~\cite{Amhis:2019ckw}. Moreover, the LFUV observables expressed in term of the ratio of branching ratios (BRs) $R_{D^{(*)}}^{\tau/\ell} = {\rm BR}(B \to D^{(*)} \tau \overline{\nu})/{\rm BR}(B \to D^{(*)} \ell \overline{\nu})$ (with $\ell=e,\mu$) and $R_{K^{(*)}}^{\mu/e}={\rm BR}(B \to K^{(*)} \mu^+\mu^-)/{\rm BR}(B \to K^{(*)} e^+e^-)$,  
are theoretically clean, i.e.,~with strongly suppressed hadronic and CKM-angle uncertainties, thus making them less vulnerable to higher-order quantum corrections~\cite{Bordone:2016gaq, Bernlochner:2017jka}. There are also other intriguing aspects of data that deserve attention. For instance, as noticed in Ref.~\cite{Altmannshofer:2020axr}, there are 11 different measurements to date on the charged-current $B$-decays $B\to D^{(*)}\ell \nu$ and $B_c\to J/\psi\ \ell \nu$  (with $\ell=e,\mu,\tau$), {\it all} of which have experimental central values above the SM central value. Similarly, both $R_K$~\cite{Aaij:2021vac} and $R_{K^{*}}$~\cite{Aaij:2017vbb} measurements from LHCb are consistently below the SM prediction of unity, which is a useful aspect of data that can be used to discriminate candidate BSM scenarios.\footnote{For example, some BSM scenarios with right-handed currents predict an anti-correlation between $R_K$ and $R_{K^*}$.} 
Also, the recently updated $R_K$ measurement from LHCb with twice as much data~\cite{Aaij:2021vac} did not  budge the central value from the old measurement (up to three decimal places, 0.846)~\cite{Aaij:2019wad}. With more statistics from LHCb and Belle-II expected in the near future, the status of the $B$-anomalies should be further clarified. 

Another long-standing anomaly that also hints at LFUV is the discrepancy between the SM and experimental values of the muon anomalous magnetic moment $(g-2)_\mu$~\cite{Jegerlehner:2009ry}. There has been an important recent development, as the first result from Fermilab $g-2$ experiment~\cite{Abi:2021gix} is found to be compatible with the old BNL result~\cite{Bennett:2006fi} to six significant figures. When combined and compared with the world-average of SM prediction~\cite{Aoyama:2020ynm}, it increases the significance of the $(g-2)_\mu$ anomaly to $4.2\sigma$~\cite{Abi:2021gix}. This does require a very accurate theoretical prediction from the SM~\cite{Aoyama:2020ynm}. Over the years a systematically improvable strategy has been developed, based on a firm theoretical foundation of dispersion relations~\cite{Lautrup:1971jf, Czarnecki:1995wq, Czarnecki:1995sz} and it has the power of being  data-driven (called the $R$-ratio method)~\cite{Colangelo:2018mtw,Benayoun:2019zwh,Hoferichter:2019mqg,Davier:2019can, Keshavarzi:2019abf}. 
In the meantime, a completely independent,  non-perturbative,  lattice simulation method, originally initiated in Ref.~\cite{Blum:2002ii}, has  significantly matured over 
the years~\cite{BMW:2017okr, Blum:2018mom,Giusti:2019xct, Shintani:2019wai,FermilabLattice:2019ugu, Gerardin:2019rua, Borsanyi:2020mff}; see Ref.~\cite{Aoyama:2020ynm} for a review. In fact, a new lattice result from the BMW collaboration~\cite{Borsanyi:2020mff} already seems to have achieved competitive accuracy and claims compatibility between its SM calculation and the experimental result at $1.5\sigma$. However, this result using a particular fermion-discretization method (``staggered fermions") disagrees with the $R$-ratio results at $\approx 3.7 \sigma$~\cite{Lehner:2020crt, Borsanyi:2020mff} and is also in $\approx 2.5 \sigma$ tension~\cite{Lehner:2020crt} with  another lattice/data (``window method”) result of RBC-UKQCD collaboration~\cite{ Blum:2018mom} using domain-wall fermions~\cite{Shamir:1992im, Furman:1994ky, Blum:1996jf, Blum:1997mz}. The domain-wall method essentially  preserves the chiral symmetry of the continuum theory at any lattice spacing and therefore, domain wall fermions  behave as continuum-like fermions so the corresponding chiral perturbation theory  is very much  continuum-like~\cite{Bernard:1985wf}. This is unlike the case of staggered fermions which involves many unphysical degrees of freedom that only decouple in the continuum limit, thus requiring a fairly cumbersome treatment of staggered chiral perturbation theory~\cite{Lee:1999zxa, Aubin:2003mg, Aubin:2003uc} for extrapolation to the continuum limit. For these reasons and many more,\footnote{For instance, another concern raised with the BMW result is that if a change in the hadronic vacuum polarization contribution brings the SM value of $(g-2)_\mu$ closer to the experimental value as claimed by the BMW collaboration, it might lead to other problems in global electroweak fit~\cite{Crivellin:2020zul, Keshavarzi:2020bfy, deRafael:2020uif, Malaescu:2020zuc} (see also the rebuttal in Ref.~\cite{Borsanyi:2020mff}). } widespread and long-standing understanding in the lattice community is that important physics results should be taken seriously {\it only after} there is consistency and agreement amongst results obtained in the continuum limit after use of as many different fermion discretization methods as possible.

Taken at face value, all three flavor anomalies, viz.~$R_{D^{(*)}}$, $R_{K^{(*)}}$,  and $(g-2)_\mu$, provide a strong   evidence for some flavor-nonuniversal BSM physics, with a combined significance of more than $5\sigma$~\cite{Altmannshofer:2020axr}. With updates from  Belle-II, LHCb and Fermilab expected soon, the chances are high that at least one of these anomalies will survive the test of time. Under such a watershed departure from the past, it is very likely that nature is also trying to address some long-standing, persistent issue(s) with the SM. One such basic concern with the SM is the fact that it is exceedingly fine-tuned, {\it i.e.~{unnatural}} due to radiative instability of the Higgs which primarily originates from the heaviness of the top quark, a member of the third generation. The LFUV  observable $R_{D^{(*)}}$ involves $b \to c \tau \nu$, where at least two fermions are from the third generation. Taking this cue from experiment and keeping  radiative stability issue in mind, along with minimality, we were originally led~\cite{Altmannshofer:2017poe} to propose a minimal $R$-parity violating (RPV) supersymmetry (SUSY) framework with the third-generation superpartners lighter than the first two (hence dubbed as `RPV3') as a compelling BSM candidate for the $R_{D^{(*)}}$ anomaly. In a follow-up work~\cite{Altmannshofer:2020axr}, we presented a benchmark point where $R_{K^{(*)}}$ and $(g-2)_
\mu$ can also be explained together with $R_{D^{(*)}}$ within the RPV3 framework. In this paper, we perform a comprehensive study of the minimal RPV3 parameter space that could simultaneously address all three flavor anomalies, in light of the recent updates and taking into account all relevant collider and flavor constraints. Let us assert that indeed there is a plethora of existing experimental constraints from $Z$, $W$, $\tau$, $D^0$, $B$, $B_s$ and $B_c$ decays, as well as $B_s-\overline{B}_s$ mixing, that have to be enforced on RPV3, thus making it rather restrictive. We then propose specific collider signatures, partly based on general crossing symmetry arguments, that could independently probe the preferred RPV3 region at the energy frontier. Such complementarity with the intensity frontier observables makes the RPV3 framework particularly predictive and testable in the near future.\footnote{Note also that there is significant motivation and rationale for this theoretical scenario from an entirely different perspective~\cite{Brust:2011tb}. Moreover, due to renormalization group (RG) evolution effects, heavy first two generations play a significant role in obtaining the observed 125 GeV Higgs mass, which makes this model less fine-tuned than the constrained Minimal Supersymmetric SM (cMSSM)~\cite{Baer:2012uy, Badziak:2012rf}.} While the RPV3 framework has many nice features, it does possess two appreciable caveats which we will briefly discuss towards the end. 

The rest of the paper is organized as follows: In Sec.~\ref{sec:frame}, we review the RPV3 framework and present all relevant expressions for the RPV3 explanation of the flavor anomalies. In Sec.~\ref{sec:scan}, we present a numerical scan of a 6-dimensional RPV3 parameter space. In Sec.~\ref{sec:BP}, we choose three benchmark points from our numerical scan and illustrate the allowed regions preferred by the anomalies. In Sec.~\ref{sec:collider}, we propose some collider signals that can be used an independent test of the anomalies. In Sec.~\ref{sec:dis}, we discuss possible ways to distinguish the RPV3 model from other BSM scenarios. Our conclusions are given in Sec.~\ref{sec:con}. In Appendix~\ref{app:constraints}, we give some details of the low-energy constraints used in our analysis.  

\section{RPV3 Framework} \label{sec:frame}
The effective number of degrees of freedom relevant to the three LFUV observables in RPV3 is of ${\cal O}(36)$ whereas in the SM it is of ${\cal O}(15)$. So the question might arise: What have we gained at the expense of doubling the degrees of freedom?  
The answer is that it has a deeper Bose-Fermi symmetry rationale, and many  associated attractive features of SUSY automatically in-built, such as radiative stability of the Higgs boson, radiative neutrino masses, radiative electroweak symmetry breaking, stability of the electroweak vacuum, gauge coupling unification, (gravitino) dark matter and  baryogenesis~\cite{Barbier:2004ez}.
Moreover, as a necessary generalization of the Yang-Mills theory~\cite{Yang:1954ek}, all the interactions allowed by the enlarged internal symmetry of the theory must be included, which readily removes the accidental flavor symmetry of the SM and leads naturally to LFUV. Our RPV3 framework also has other non-trivial features that are consistent with the experimental observations, such as having both $R_K$ and $R_{K^*}$ less than one and the $D^*$ and tau polarizations essentially the same as in the SM, both of which come automatically because RPV3 subsumes the chiral gauge couplings of the SM.

It is worth noting here that the  semileptonic $B$-meson decays in question involve interactions of a bottom quark, a member of the third-generation, in terms of either $b\to c\ell^- \overline{\nu}$ or $b\to s\ell^+\ell^-$ transitions. Analogous semileptonic decays of charmed mesons $D\to X\ell^+\nu$ ($X=\pi, K, \omega,  \eta, \rho$)~\cite{Ablikim:2018frk, Ablikim:2018evp, BESIII:2020dbj, BESIII:2020dtz, Ablikim:2021gct},  all involving $c\to d\ell^+ \nu$ transition, as well as the ratio of the rates of  leptonic kaon decays $K^\pm\to e^\pm \nu$ and $K^\pm \to \mu^\pm \nu$~\cite{Ambrosino:2009aa, Lazzeroni:2012cx}, and of $\Lambda$-baryon decays $\Lambda\to pe^-\overline{\nu}$ and $\Lambda\to p\mu^-\overline{\nu}$~\cite{BESIII:2021ynj} are all in complete agreement with the SM. Therefore, it is conceivable that the third-generation fermions are special in the SM, and in the same vein, we take the third-generation sfermions to be special in RPV3 (similar to the `natural SUSY' hypothesis~\cite{Brust:2011tb, Papucci:2011wy}). 

The $R_{D^{(*)}}$ anomaly can be  accommodated in RPV3 at tree-level via the $LQD$ interactions~\cite{Altmannshofer:2017poe, Deshpande:2012rr,  Zhu:2016xdg,Deshpand:2016cpw,Trifinopoulos:2018rna, Hu:2018lmk,Trifinopoulos:2019lyo, Wang:2019trs}:
\begin{align} 
& {\cal L}_{LQD} \ = \   \lambda^\prime_{ijk}\big[\widetilde{\nu}_{iL}\overline{d}_{kR}d_{jL}+\widetilde{d}_{jL}\overline{d}_{kR}\nu_{iL}+\widetilde{d}^*_{kR}\overline{\nu}^c_{iL}d_{jL}\nonumber \\
& \quad  - \widetilde{e}_{iL}\overline{d}_{kR}u_{jL}-\widetilde{u}_{jL}\overline{d}_{kR}e_{iL}-\widetilde{d}^*_{kR}\overline{e}^c_{iL}u_{jL}\big]+{\rm H.c.}
\label{Eq.lambda_prime}
\end{align}
Similarly, the $R_{K^{(*)}}$ anomaly can be explained via both tree and loop-level $LQD$ interactions alone or together with $LLE$ interactions~\cite{Biswas:2014gga, Deshpand:2016cpw, Das:2017kfo,  Earl:2018snx,Trifinopoulos:2018rna,Trifinopoulos:2019lyo, Hu:2019ahp, Zheng:2021wnu}: 
\begin{align}
{\cal L}_{LLE} \ = \ & \frac{1}{2}\lambda_{ijk}\big[ \widetilde{\nu}_{iL} \overline{e}_{kR} e_{jL} +\widetilde{e}_{jL} \overline{e}_{kR}\nu_{iL} +\widetilde{e}_{kR}^{*} \overline{\nu}_{iL}^c e_{jL} \nonumber \\
& \qquad - (i\leftrightarrow j) \big]+{\rm H.c.}
\label{Eq.RPVLLE}
\end{align}
The muon $g-2$ gets additional contributions from both $LQD$ and $LLE$ terms~\cite{Kim:2001se}, but as we will see later, the $LLE$ contribution is more relevant for our parameter space of interest.\footnote{The simultaneous presence of $\lambda$ and $\lambda^\prime$ couplings is consistent with proton decay constraints, as long as the relevant  $\lambda^{\prime\prime}$ ($UDD$-type) couplings are sufficiently suppressed~\cite{Barbier:2004ez}, which can be done using a baryon triality~\cite{Ibanez:1991hv, Ibanez:1991pr}.}  

Out of the $3^3=27$ independent RPV couplings $\lambda^\prime_{ijk}$ in Eq.~(\ref{Eq.lambda_prime}) and the $3^2=9$ independent $\lambda_{ijk}$ (since it is antisymmetric in the first two indices, i.e.~$\lambda_{ijk}=-\lambda_{jik}$) in Eq.~(\ref{Eq.RPVLLE}), we only consider those involving third-generation sfermions in our RPV3 framework. In what follows, we calculate the RPV3 contributions to the flavor anomalies. 

\subsection{$R_D$ and $R_{D^{(*)}}$}
\begin{figure}[b!]
		\centering
		\includegraphics[width=0.49\linewidth]{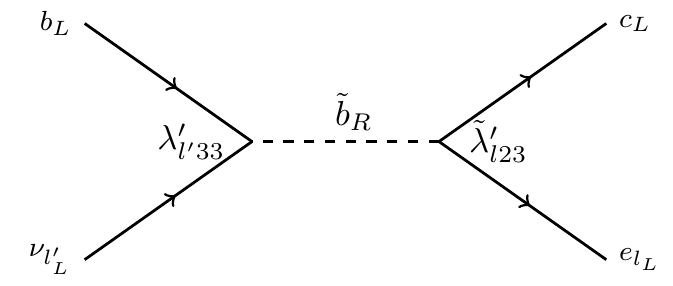}
	\caption{RPV3 contribution to $R_{D^{(*)}}$ via sbottom exchange involving $\lambda'$ couplings. Here $\widetilde{\lambda'}_{ijk}$ is defined as $\lambda'_{ilk}V_{jl}$ (with $V_{jl}$ being the CKM matrix elements). The complete set of diagrams can be found in Ref.~\cite{Altmannshofer:2020axr}. }
	\label{fig:RDdiagram}
\end{figure}
The $b\to c\ell \nu$ transition relevant for the $R_{D^{(*)}}$ anomaly gets a BSM contribution at tree level from the $LQD$ interactions via the right-handed sbottom ($\widetilde{b}_R$) exchange, as shown in Fig.~\ref{fig:RDdiagram}. It gives rise to a SM-like effective Hamiltonian 
\begin{align}
{\cal H}_{\text{eff}}^{b\to c \ell \nu}  \ = \  & \frac{4G_F}{\sqrt{2}}V_{cb}\left(  1 + C_{V_L} \right) {\cal O}_{V_L} + \text{H.c.} 
\label{eq:effH}
\end{align}
(where $G_F$ is the Fermi constant and $V_{cb}$ is the (2,3) CKM element) 
with the operator ${\cal O}_{V_L}=(\overline{c}_L\gamma^\mu b_L)(\overline{\ell}_L\gamma_\mu \nu_{\ell L})$ with a corresponding coefficient $C_{V_L}\simeq 0.09$, as  preferred by the $q^2$ and $D^*$ polarization data~\cite{Murgui:2019czp}. 
We then have~\cite{Trifinopoulos:2018rna} 
\begin{align} \label{eq:RDLHCb}
 & \frac{R_D}{R_D^\text{SM}} \ = \  \frac{R_{D^*}}{R_{D^*}^\text{SM}} = \frac{|\Delta^c_{31}|^2+|\Delta^c_{32}|^2+|1+\Delta^c_{33}|^2}
{|\Delta^c_{21}|^2+|1+\Delta^c_{22}|^2+|\Delta^c_{23}|^2} ~, \\
%\displaybreak
%\end{equation}
%where %
%\begin{align}\label{eq:delta}
& {\rm with}~\Delta_{ll'}^c \ = \ \frac{v^2}{4 m^2_{\widetilde b_R}} \lambda_{l'33}^\prime \left( \lambda_{l33}^\prime + \lambda_{l23}^\prime \frac{V_{cs}}{V_{cb}} + \lambda_{l13}^\prime \frac{V_{cd}}{V_{cb}} \right)  \nonumber
%+& \frac{v^2}{\blue{4} m^2_{\widetilde {\tau}_L}} \lambda_{l'3l} \left( \lambda_{333}^\prime + \lambda_{323}^\prime \frac{V_{cs}}{V_{cb}} + \lambda_{313}^\prime \frac{V_{cd}}{V_{cb}} \right)
~,
\end{align}
$v = (\sqrt 2 G_F)^{-1/2}$ being the electroweak scale. The $R_{D^{(*)}}$ anomaly can be explained for the ratio in Eq.~\eqref{eq:RDLHCb} being $1.15 \pm 0.04$~\cite{Altmannshofer:2020axr}, 
%
%\begin{equation}
% \frac{R_D}{R_D^\text{SM}} \ = \  \frac{R_{D^*}}{R_{D^*}^\text{SM}} \ = \ 1.15 \pm 0.04 ~,
%\end{equation}
%
which dictates the $R_{D^{(*)}}$-allowed parameter space in the $(m_{\widetilde{b}_R},\lambda^\prime_{lk3})$ plane. 

From Eq.~\eqref{Eq.lambda_prime} we see that for the $LQD$ interactions in RPV3, the dimension-six effective interaction for the semileptonic $B\to D^{(*)}$ decays is essentially identical to the $(V - A)\times (V -A)$ structure of the SM effective Hamiltonian (after the appropriate Fierz transformation) with the difference being just in the overall coefficient. Therefore, the fact that the experimentally observed $q^2$ distribution and the $D^*$ and $\tau$ polarizations prefer the ${\cal O}_{V_L}$ operator~\cite{Murgui:2019czp} is consistent with our RPV3 scenario.    

In presence of $LLE$ interactions, there is another contribution to $R_{D^{(*)}}$ from RPV3 with left-handed stau exchange~\cite{Altmannshofer:2020axr}; however, this involves right-handed bottom and charged-lepton, and the corresponding effective operator ${\cal O}_{V_R}$ does not provide the best-fit to the $b\to c\ell \nu$ observables~\cite{Murgui:2019czp}. Therefore, we choose the coupling $\lambda^\prime_{323}=0$ so that the stau channel is not relevant. Similarly, for a light neutralino $\widetilde{\chi}$, there are additional contributions involving $B\to D^{(*)}\ell\widetilde{\chi}$, which however turn out to be sub-dominant~\cite{Altmannshofer:2020axr}. 

It is also important to note that in the MSSM with two Higgs doublets, there is a standard $R$-parity conserving (RPC) contribution to $b\to c\ell\nu$ due to charged Higgs exchange. But this goes in the wrong direction and is much smaller for $R_{D^*}$~\cite{BaBar:2013mob, Prim:2019hyn}. Moreover, this is in tension with LHC mono-$\tau$ data~\cite{Greljo:2018tzh} and also induces a large ${\rm BR}(B_c\to \tau \nu)>50\%$ which is problematic~\cite{Alonso:2016oyd, Akeroyd:2017mhr, Aebischer:2021ilm} (see however Ref.~\cite{Iguro:2022uzz}). Therefore, one must resort to the RPV interactions given above to explain the $b\to c\ell\nu$ anomalies within a SUSY framework.  

\subsection{$R_{K}$ and $R_{K^{(*)}}$}

As for the $R_{K^{(*)}}$ anomaly involving $b\to s\ell^+\ell^-$ transitions, simultaneous RPV contributions to electron and muon final states would be strongly constrained by lepton flavor violating (LFV) searches like $\mu\to e\gamma$. Therefore, we only consider corrections to the muonic channel, as preferred by recent global fits~\cite{Altmannshofer:2021qrr}. The relevant effective Hamiltonian is  
\begin{equation}
    \mathcal H_\text{eff}^{b\to s\ell\ell}  =  - \frac{4 G_F}{\sqrt{2}} V_{ts}^* V_{tb} \frac{e^2}{16\pi^2} \sum_{i = 9,10}\left[ C_i^\ell Q_i^\ell + C_i^{\prime\ell} Q_i^{\prime\ell} \right]  
    \label{eq:Heff}
\end{equation}
with the operators
%
%\begin{align}
$ Q_9^\ell  =  (\overline{ s} \gamma_\alpha P_L b)(\overline{ \ell} \gamma^\alpha \ell)$, 
$Q_{10}^\ell   =  (\overline{ s} \gamma_\alpha P_L b)(\overline{ \ell} \gamma^\alpha \gamma_5 \ell)$, 
and $Q_{9,10}^\prime$ are obtained from $Q_{9,10}$ by replacing $P_L \to P_R$. Global fits of all relevant data, including anugular observables or absolute rate for $B\to K^{(*)}\mu^+\mu^-$ and also rate for $B_s\to \phi \mu^+\mu^-$, prefer the Wilson coefficients $C_9^\mu  =  -C_{10}^\mu  =  -0.35 \pm 0.08$~\cite{Altmannshofer:2021qrr},  
%\begin{align}
%C_9^\mu \ = \ -C_{10}^\mu \ = \ -0.35 \pm 0.08 ~,
%\label{eq:global}
%\end{align}
whereas $C_9^{\prime\mu}$ and  $C_{10}^{\prime\mu}$ are compatible with zero at $2\sigma$ level. 
%\begin{align}
%C_9^{\prime\mu} \ = \ -0.32^{+0.16}_{-0.17} \, , \quad C_{10}^{\prime\mu} \ = \ 0.06\pm 0.12 \, .
%\label{eq:global2}
%\end{align}

In the RPV3 scenario, new contributions to $b \to s \ell \ell$ transitions arise both at  tree and loop levels.
Tree-level exchange of stops gives contributions to the wrong chirality Wilson coefficients~\cite{Das:2017kfo}, which  
%:
%
%\begin{equation}
% C_9^{\prime\mu} \ = \  - C_{10}^{\prime^\mu} = - \frac{v^2}{2m_{\widetilde t_L}^2} \frac{\pi}{\alpha_\text{em}} \frac{\lambda^\prime_{233} \lambda^\prime_{232}}{V_{tb} V_{ts}^*} ~,
% \label{eq:C9pC10p}
%\end{equation}
%
%where $\alpha_\text{em}$ is the fine structure constant. 
%The global-fit results~\eqref{eq:global2} 
can be translated into an approximate $3\sigma$ confidence level (CL) upper bound on 
\begin{equation}
\left| \lambda^\prime_{233} \lambda^\prime_{232} \right| \ \lesssim  \ 10^{-3} \times \left( \frac{m_{\widetilde t_L}}{1\,\text{TeV}} \right)^2 ~. \label{eq:primebound}
\end{equation}
This can be satisfied by either making the stop relatively heavier, or by setting one of $\lambda^\prime_{23k}$ (with $k=2$ or 3) small. 
%In the benchmark point discussed below, we will choose $\lambda^\prime_{232}=0$ to satisfy this constraint. 

%%%%%%%%%%%%%
\begin{figure}[t!]
		\centering
		\includegraphics[width=0.49\linewidth]{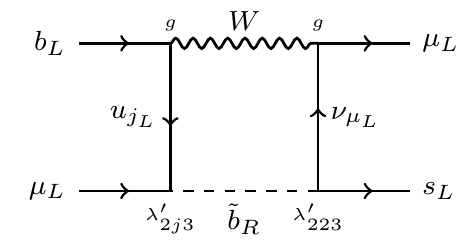}  
		\includegraphics[width=0.49\linewidth]{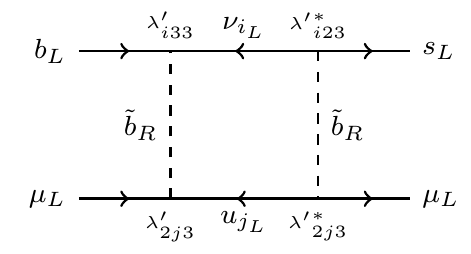}  
	\caption{Representative box diagrams for the dominant RPV3 contributions to $b \to s \mu^+\mu^-$ in our scenario. The complete set of diagrams can be found in Ref.~\cite{Altmannshofer:2020axr}. }
	\label{fig:C9C10diagrams}
\end{figure}
%%%%%%%%%%%%%%%%%%%%%%%%%%%%%%

At one-loop level, there are new contributions to $b\to s\ell\ell$ with sbottoms, 
stops, staus and sneutrinos in the loop~\cite{Das:2017kfo,Earl:2018snx,Trifinopoulos:2018rna, Altmannshofer:2020axr}. For the minimal RPV3 case considered here, the only relevant diagrams are those shown in Fig.~\ref{fig:C9C10diagrams} and the resulting Wilson coefficients (after taking into account all possible combinations of box diagrams) are given by
\begin{eqnarray} \label{eq:C9C10} 
C_9^\mu   =   - C_{10}^\mu  =   \frac{m_t^2}{m_{\widetilde b_R}^2} \frac{|\lambda^\prime_{233}|^2}{16\pi\alpha_\text{em}} - \frac{v^2}{m_{\widetilde b_R}^2} \frac{X_{bs} X_{\mu\mu}}{64\pi \alpha_{\rm em} V_{tb} V_{ts}^*} ,
%\nonumber \\
%&& -\frac{v^2}{16 ( m_{\widetilde t_L}^2-m_{\widetilde \nu_\tau}^2)} \log\left(\frac{m_{\widetilde t_L}^2}{m_{\widetilde \nu_\tau}^2}\right) \frac{X_{b\mu} X_{s\mu}}{e^2 V_{tb} V_{ts}^*} \nonumber\\
%&& -\frac{v^2}{16 ( m_{\widetilde b_R}^2-m_{\widetilde \tau_R}^2)} \log\left(\frac{m_{\widetilde b_R}^2}{m_{\widetilde \tau_R}^2}\right) \frac{\widetilde X_{b\mu} \widetilde X_{s\mu}}{e^2 V_{tb} V_{ts}^*} \nonumber\\
%&& -\frac{v^2}{16 m_{\widetilde \nu_\tau}^2} \frac{\widetilde X_{bs} \widetilde X_{\mu\mu}}{e^2 V_{tb} V_{ts}^*} ~,
\end{eqnarray}
where $X_{bs}=\sum_{i=1}^3\lambda^\prime_{i33} \lambda^\prime_{i23}$ and $X_{\mu\mu}=\sum_{j=1}^3|\lambda^\prime_{2j3}|^2$.
%the $X$ and $\widetilde X$ factors are different combinations of RPV couplings:
%\begin{eqnarray}
%X_{bs} & \ = \ & \lambda^\prime_{133} \lambda^\prime_{123} + \lambda^\prime_{233} \lambda^\prime_{223} + \lambda^\prime_{333} \lambda^\prime_{323} ~, \nonumber \\
%\widetilde X_{bs} & \ = \ & \lambda^\prime_{331} \lambda^\prime_{321} +\lambda^\prime_{332} \lambda^\prime_{322} +\lambda^\prime_{333} \lambda^\prime_{323} ~, \nonumber \\
%X_{\mu\mu} & \ = \ & |\lambda^\prime_{213}|^2 + |\lambda^\prime_{223}|^2 + |\lambda^\prime_{233}|^2  ~, 
%\nonumber \\
%\widetilde X_{\mu\mu} & \ = \ & |\lambda_{231}|^2 + |\lambda_{232}|^2 + |\lambda_{233}|^2  ~, \nonumber \\
%X_{b\mu} & \ = \ & \lambda^\prime_{331} \lambda^\prime_{231} + \lambda^\prime_{332} \lambda^\prime_{232} + \lambda^\prime_{333} \lambda^\prime_{233} ~, \nonumber \\
%X_{s\mu} & \ = \ & \lambda^\prime_{321} \lambda^\prime_{231} +\lambda^\prime_{322} \lambda^\prime_{232} +\lambda^\prime_{323} \lambda^\prime_{233}~, \nonumber \\
%\widetilde X_{b\mu} & \ = \ & \lambda^\prime_{133} \lambda_{123} + \lambda^\prime_{333} \lambda_{323} ~, \nonumber \\
%\widetilde X_{s\mu} & \ = \ & \lambda^\prime_{123} \lambda_{123} +\lambda^\prime_{323} \lambda_{323}~. 
%\end{eqnarray}
 Requiring Eq.~\eqref{eq:C9C10} to match the global-fit result yields the $R_{K^{(*)}}$-allowed parameter space in the $(m_{\widetilde{b}_R},\lambda^\prime_{lk3})$ plane. Note that the correlation between $R_K$ and $R_{K^{(*)}}$, i.e. both going in the same direction is automatically obtained in the RPV3 setup because of the same underlying gauge structure as in the SM.  

It is important to note here that in the MSSM with RPC couplings only, the only way to get lepton-flavor non-universal contribution to $b\to s\ell\ell$ is through box diagrams with light winos (or binos) and large non-universality in slepton masses~\cite{Altmannshofer:2013foa, Altmannshofer:2014rta}. However, in order to get the required $C_9^\mu\sim -0.35$, one needs an extremely light spectrum of winos and smuons around 100 GeV and sbottoms around 500 GeV, which are ruled out by the LHC data~\cite{ParticleDataGroup:2020ssz}. Therefore, just like in the case of $R_{D^{(*)}}$, the explanation of the $R_{K^{(*)}}$ anomaly within SUSY requires one to invoke RPV interactions.

\subsection{$(g-2)_\mu$}
\begin{figure}[t!]
		\centering
		\includegraphics[width=0.49\linewidth]{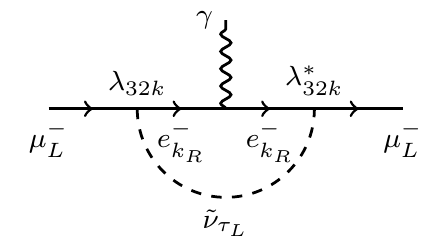}  
		\includegraphics[width=0.49\linewidth]{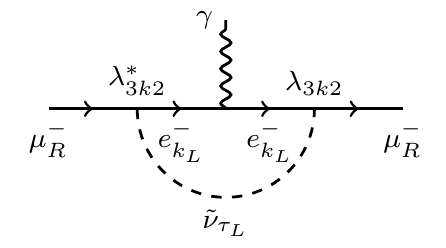} 
		\includegraphics[width=0.49\linewidth]{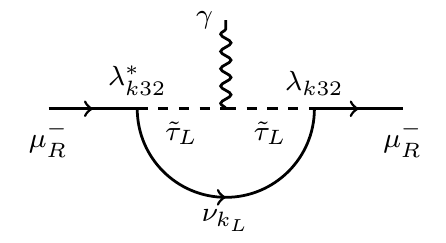}  
		\includegraphics[width=0.49\linewidth]{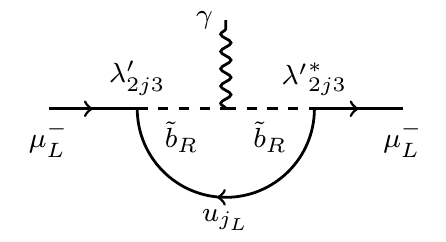}  
		\includegraphics[width=0.49\linewidth]{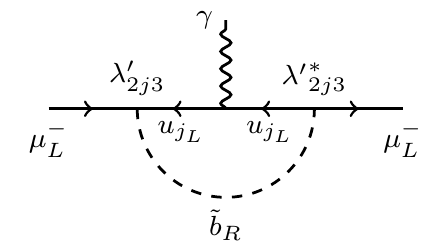}  
	\caption{Relevant contribution to the $(g-2)_{\mu}$ from $\lambda$ and $\lambda'$ couplings in our RPV3 scenario. The complete set of diagrams can be found in Ref.~\cite{Altmannshofer:2020axr}. 
	}
	\label{fig:muongm2}
\end{figure}
The RPV3 contributions to $(g-2)_{\mu}$ can arise from both $\lambda$ and $\lambda^\prime$ couplings~\cite{Kim:2001se}, as shown in Fig.~\ref{fig:muongm2}:
\begin{align}
    \Delta a_{\mu} & \ = \ \frac{m_{\mu}^2}{96\pi^2}\sum_{k=1}^3 \left(\frac{2(|\lambda_{32k}|^2+|\lambda_{3k2}|^2)}{m^2_{\widetilde{\nu}_{\tau}}}\right.\nonumber \\
    & \qquad \left.-\frac{|\lambda_{3k2}|^2}{m^2_{\widetilde{\tau}_{L}}}
    -\frac{|\lambda_{k23}|^2}{m^2_{\widetilde{\tau}_{R}}}
    +\frac{3|\lambda^\prime_{2k3}|^2}{m_{\widetilde{b}_R}^2} \right)
    \label{eq:gm2l} \, ,
\end{align}
which should be compared with the observed discrepancy of $\Delta a_\mu^{\rm obs}=(251 \pm 59)\times  10^{-11}$~\cite{Abi:2021gix}. Note that the stau contributions are of the wrong sign,  and therefore, are required to be sub-dominant by making the staus relatively heavier. As we will see later, the $\lambda$-contribution to $(g-2)_\mu$ from sneutrinos is dominant over the $\lambda'$-contribution from sbottoms in our RPV3 scenario. There are other diagrams, as shown in Ref.~\cite{Altmannshofer:2020axr}, which are not relevant to our discussion. For instance, the stop-mediated diagrams cancel, so are not shown in Fig.~\ref{fig:muongm2}. Similarly, the diagram with right-handed stau mediator does not contribute because one of the corresponding couplings is assumed to be zero here. 

There are additional RPC SUSY contributions to $\Delta a_\mu$ involving  
smuons and muon sneutrinos~\cite{Moroi:1995yh, Czarnecki:2001pv, Baum:2021qzx}. However, since these second-generation sfermions are 
decoupled from the low-energy theory in RPV3, we only focus on the RPV contributions. 

\section{Numerical Scan} \label{sec:scan} 
Our aim is to find the {\it minimum} set of RPV3 model parameters that could simultaneously address all three flavor anomalies,  while being consistent with all other low-energy flavor constraints (for a detailed discussion and explicit expressions, see Appendix~\ref{app:constraints} and Ref.~\cite{Altmannshofer:2020axr}\footnote{We found that the parameter setup of Case 3 in Ref.~\cite{Altmannshofer:2020axr} which could explain all flavor anomalies is actually problematic for the LFV decays $\mu\to e\gamma$ and $B_s\to e\mu$. This prompted us to look for new solutions in this work.}) and the high-energy LHC data, while at the same time  giving rise to potentially observable collider signals as an independent test of the anomalies. To this effect, we choose to work with the following 6-dimensional parameter space that turns out to be the most important for pursuing collider implications:  
\begin{align}
    \{\lambda_{232}, \lambda_{233}^\prime, \lambda_{223}^\prime, \lambda_{232}^\prime,  m_{\widetilde{b}_{R}}, %m_{\widetilde{b}_{L}},
    m_{\widetilde{\nu}_\tau} %m_{\widetilde{\tau}_{L}}
    \}
    \label{eq:BP}
\end{align}
and drop the other couplings and masses from our discussion (unless otherwise specified).\footnote{Because the RG evolution of any RPV coupling is always proportional to the coupling itself (at least up to two-loop level)~\cite{Allanach:1999mh}, the couplings set to zero at the input scale remain zero at all scales under the RG flow. }
As for the above choice of our couplings, note that once we choose a nonzero $\lambda_{232}=-\lambda_{322}$ to explain the $(g-2)_\mu$ anomaly [cf.~Eq.~\eqref{eq:gm2l}], 
other relevant $\lambda_{3ij}$ couplings cannot be large at the same time due to the constraints from LFV decays $\tau^-\to \mu^+\mu^-\mu^-$ and $\mu\to e\gamma$. Similarly, only some of the $\lambda^\prime_{2ij}$'s are allowed to be large at the same time as $\lambda_{232}$; if instead we chose $\lambda^\prime_{3ij}$, for instance, combined with $\lambda_{232}$ and light tau-sneutrino propagator, this will lead to strong tree-level meson decays $\overline{d}_id_j\to \mu^+\mu^-$. Note that due to our choice of couplings (i.e. $\lambda'_{3ij}=0$), we do not have the single production of tau-sneutrino at the LHC, and therefore, only discuss their pair production in the main text.   

As for omitting the remaining third-generation sfermion masses from Eq.~\eqref{eq:BP}, the right-stau mass $m_{\widetilde{\tau}_R}$ is irrelevant, because it only enters in the $(g-2)_\mu$ expression~\eqref{eq:gm2l}, 
but with the coupling $\lambda_{k23}$, which is set to zero for our benchmark points. The left-stau term in Eq.~\eqref{eq:gm2l} 
has a negative contribution and its effect can be ignored for $m_{\widetilde{\tau}_{L}} \gtrsim {\cal O}(2\ \mathrm{TeV})$. For concreteness, we will just set $m_{\widetilde{\tau}_{L}}=4$ TeV in the following analysis. Similarly, the left-stop mass $m_{\widetilde{t}_L}$ only influences the $C_9'$ and $C_{10}'$ [cf. Eq.~\eqref{eq:primebound}] and the $B_s\to \mu^+\mu^-$ constraint, if both $\lambda^\prime_{233}$ and $\lambda^\prime_{232}$ are large at the same time. When this is the case, we can make $m_{\widetilde{t}_L}$ appropriately heavier using Eq.~\eqref{eq:primebound} 
without affecting any other observable; therefore, we do not include $m_{\widetilde{t}_L}$ in Eq.~\eqref{eq:BP}. Finally, the left-sbottom mass $m_{\widetilde{b}_L}$ does not influence the anomaly observables, but is only relevant for constraints like $b\to s\gamma$, $B\to K\nu\overline{\nu}$ and $B_s$-$\overline{B}_s$; so wherever  applicable, we will just set $m_{\widetilde{b}_L}=m_{\widetilde{b}_R}$ for simplicity.   

%Note that we are primarily focusing on the $>3\sigma$ LFUV anomalies only as they are theoretically cleaner and also their chances of survival are rather high. We have not included other possible indications of deviations from the SM, %such as angular observables
%or absolute rate for $B \to K^{(*)} \mu^+ \mu^-$~\cite{Aebischer:2019mlg} and also rate for $B_s \to \phi \mu^+ \mu^-$~\cite{LHCb:2015wdu} as in these 
%cases there can be non-perturbative 
%contributions from non-local operators especially in the region of low $q^2$ that are not under full theoretical control yet. 
%Similarly, we do not include the ANITA anomaly~\cite{} because ... 
%and also CKM anomaly.
%Similarly, we do not include the $(g-2)_e$ anomaly, because of a $>5\sigma$ discrepancy between the Cs~\cite{Hanneke:2010au} and Rb~\cite{Morel:2020dww} measurements of the fine-structure constant, so it is not clear which of these results should be used for comparison of the experimental value with the SM prediction~\cite{Aoyama:2014sxa} for $(g-2)_e$. 

%We will make some further assumptions (see {\it Supplemental Material}) to reduce the number of free parameters for our benchmark point discussed below. 

\begin{figure*}[t!]
    \centering
    \subfigure[]{
    \includegraphics[width=0.323\textwidth]{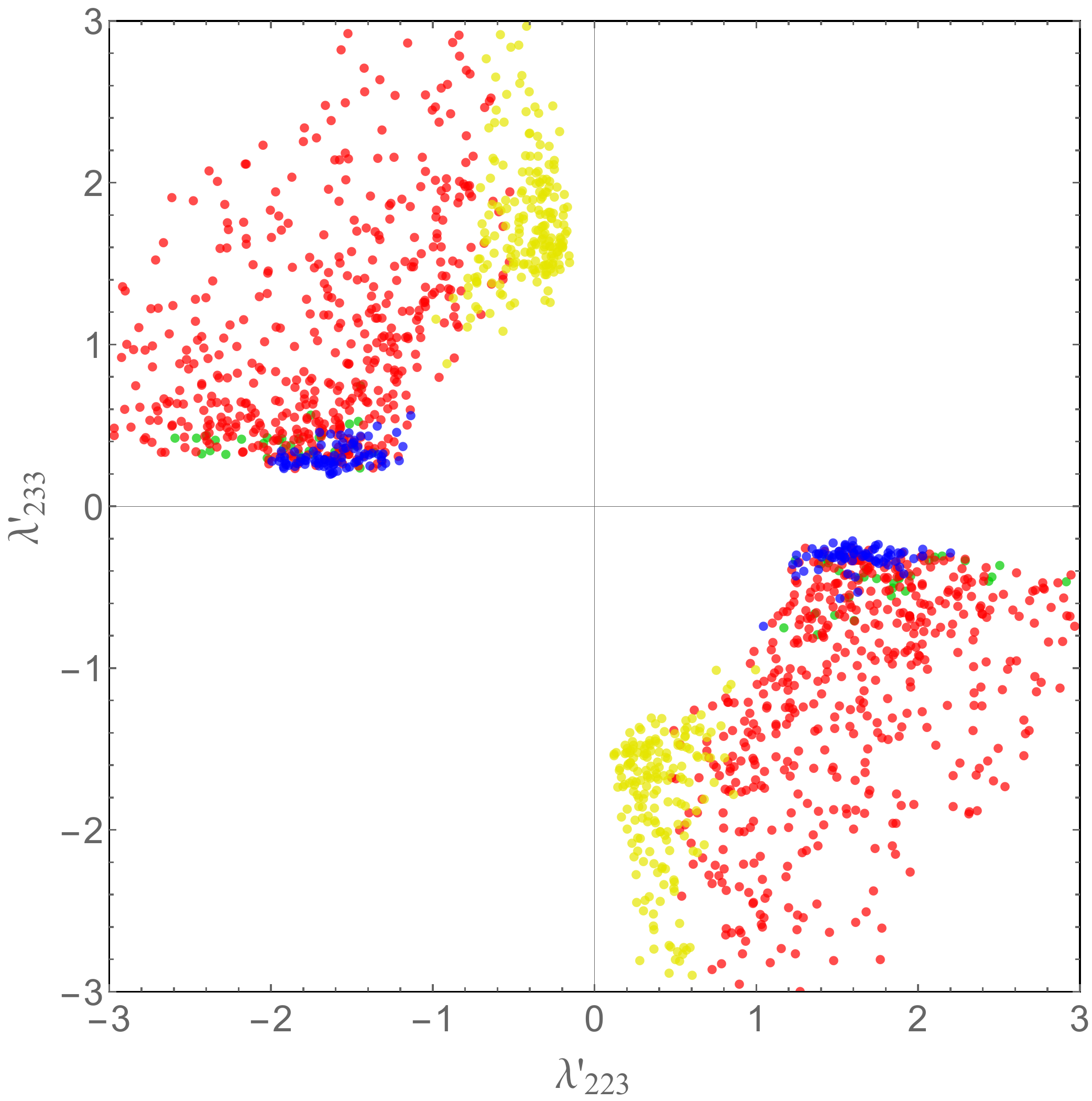}}
   \subfigure[]{ \includegraphics[width=0.323\textwidth]{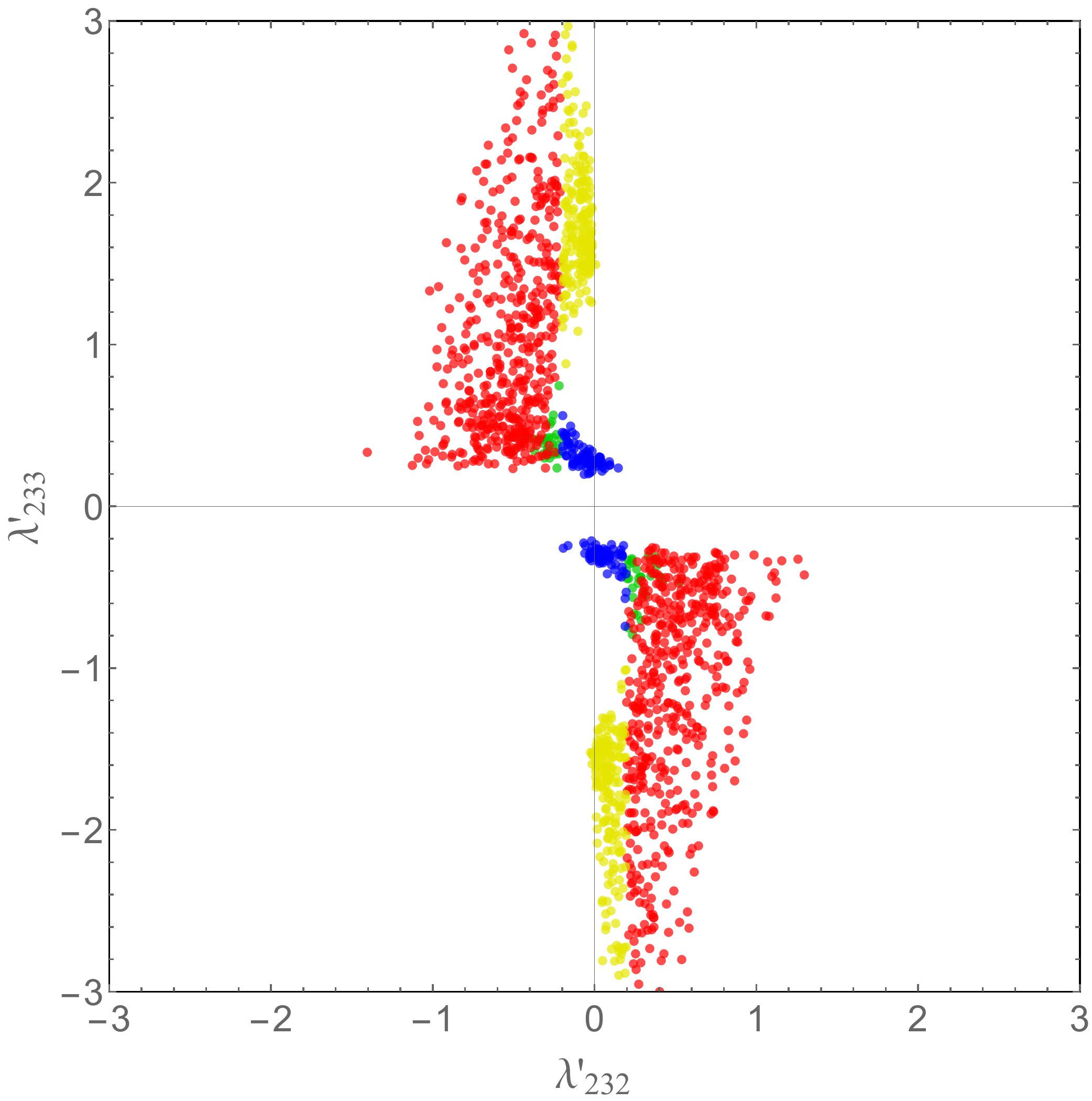}}
   \subfigure[]{ \includegraphics[width=0.323\textwidth]{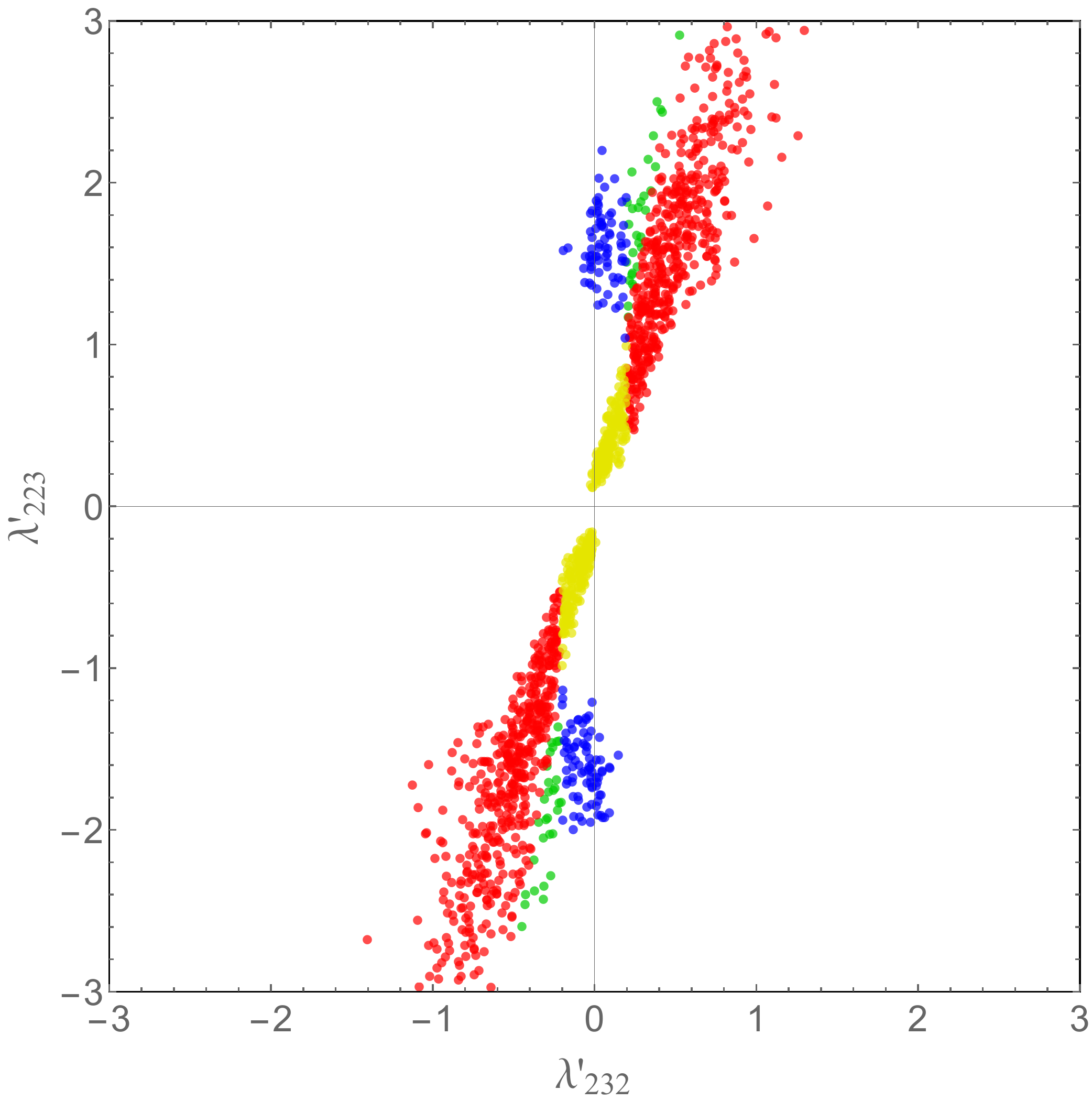}}
      \\
       \subfigure[]{\includegraphics[width=0.325\textwidth]{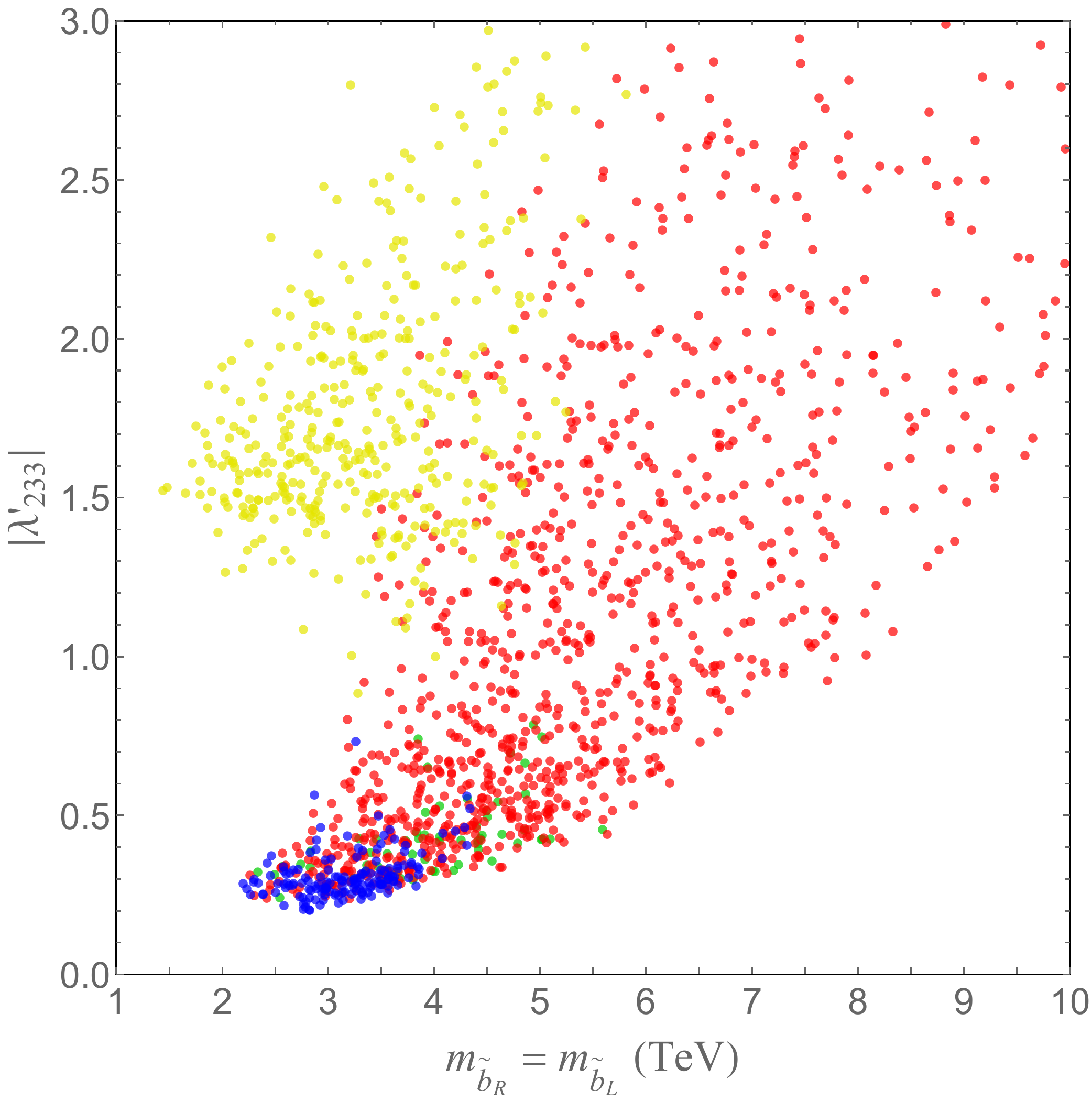}}
       \subfigure[]{\includegraphics[width=0.325\textwidth]{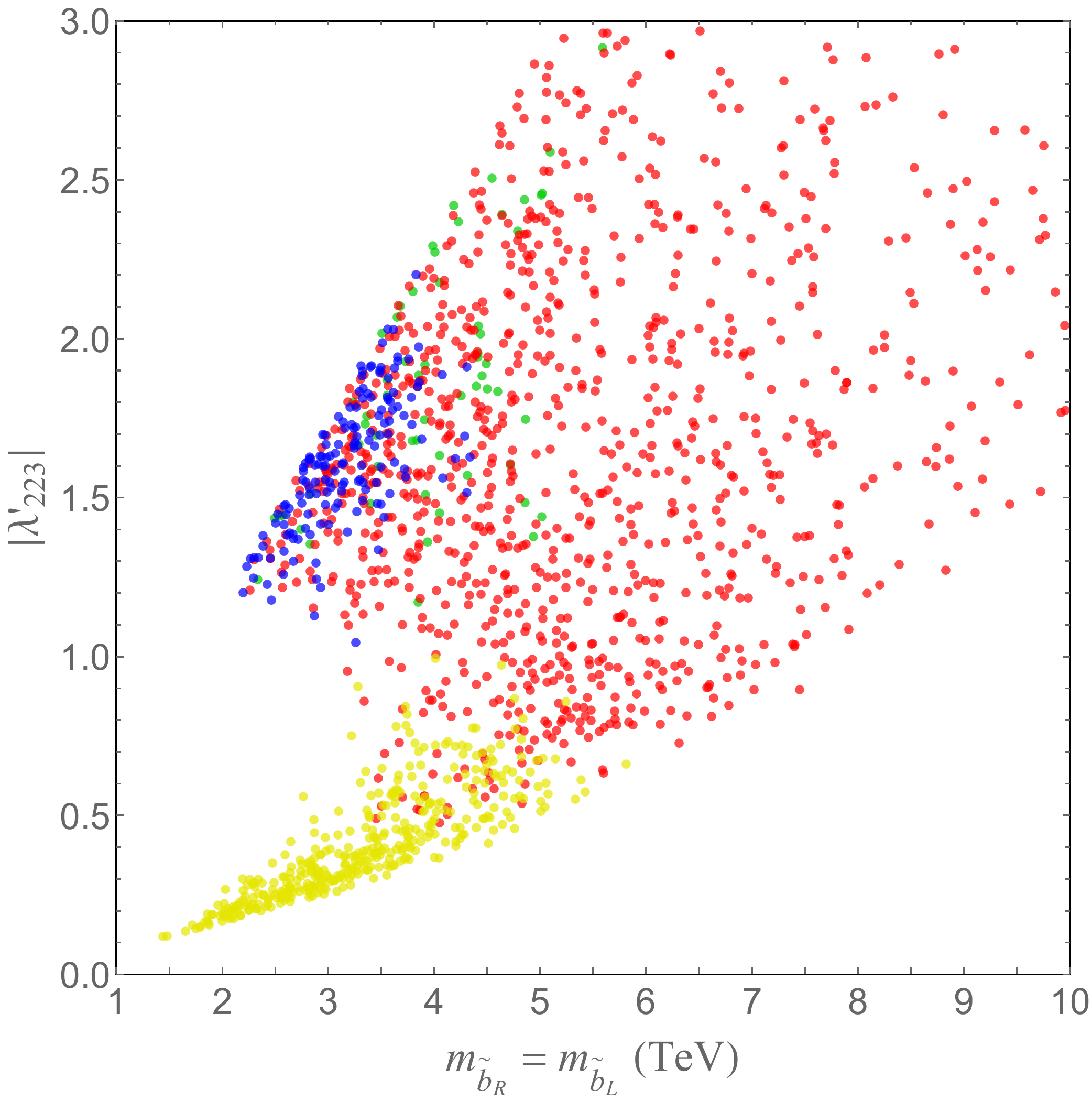}}
       \subfigure[]{\includegraphics[width=0.325\textwidth]{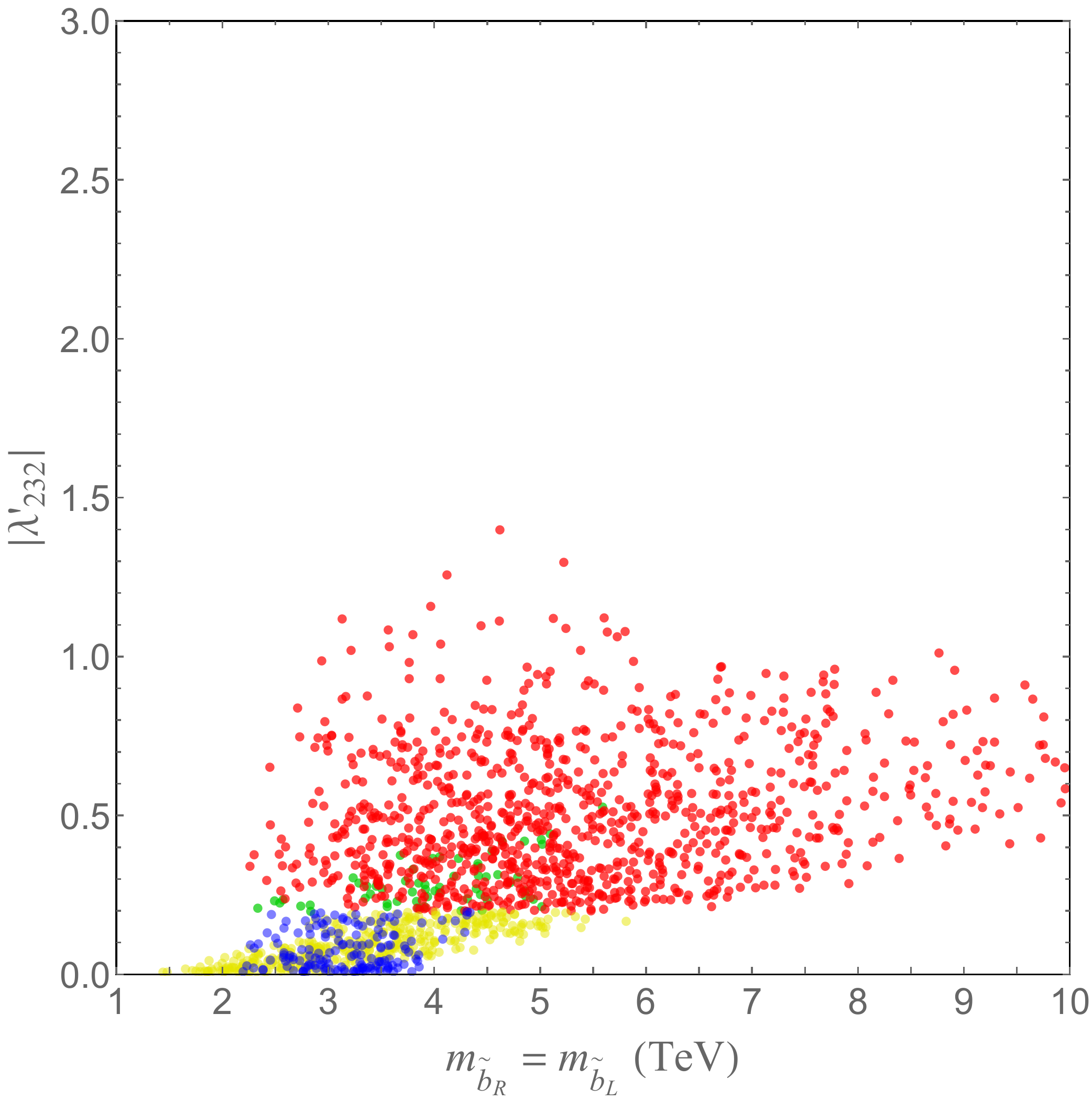}}\\
       \subfigure[]{\includegraphics[width=0.325\textwidth]{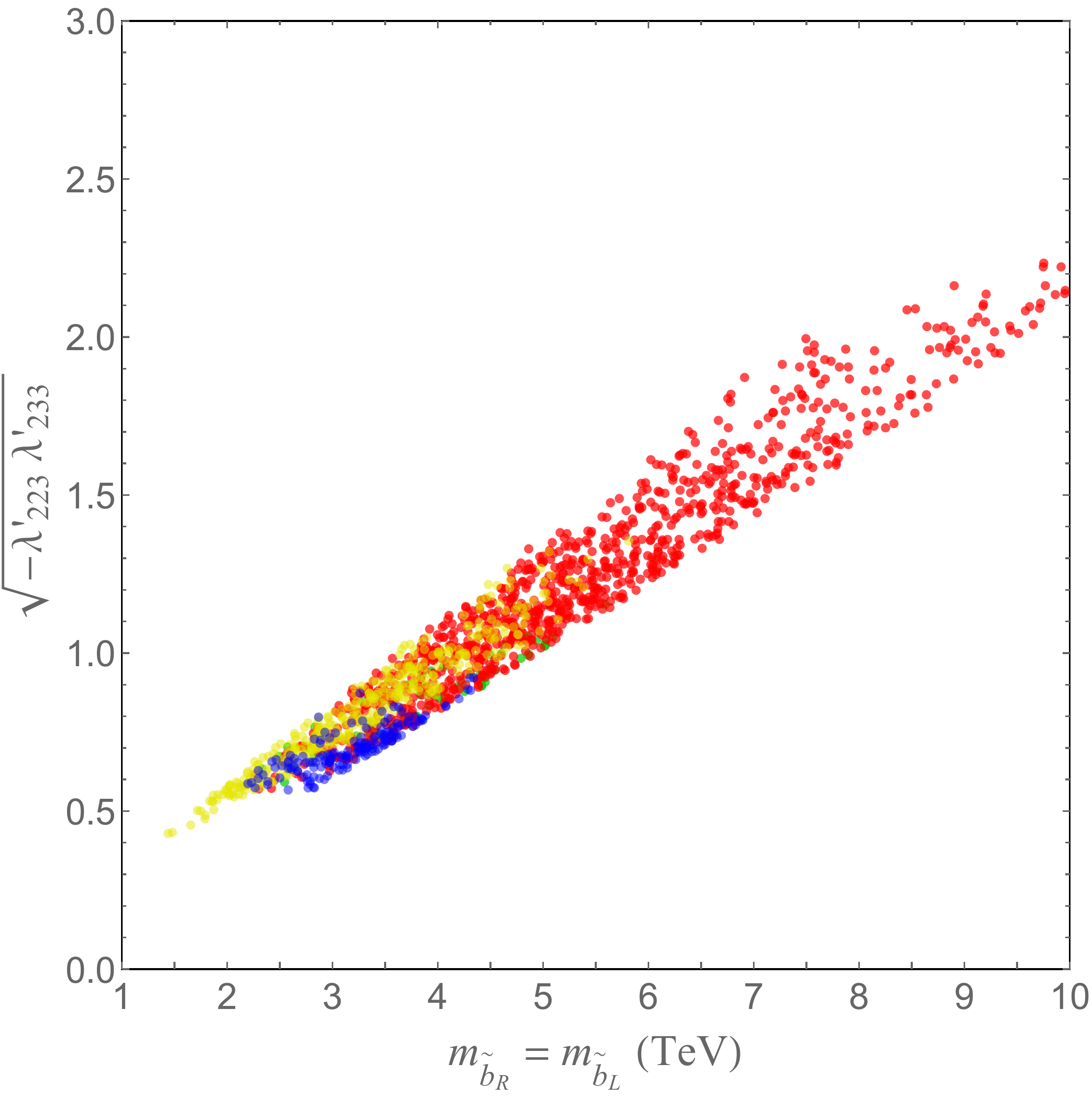}}
       \subfigure[]{\includegraphics[width=0.325\textwidth]{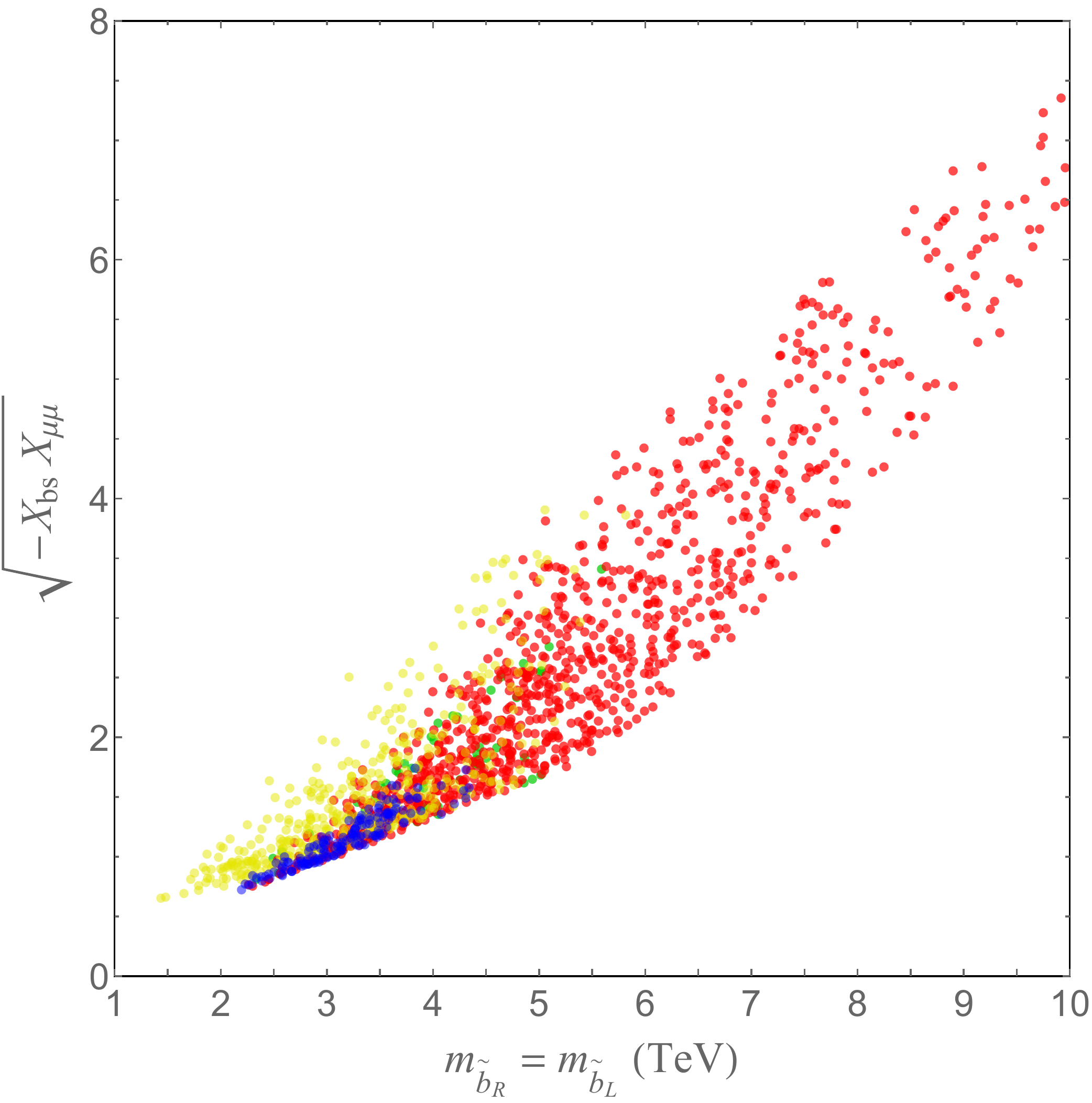}}
\subfigure[]{\includegraphics[width=0.325\textwidth]{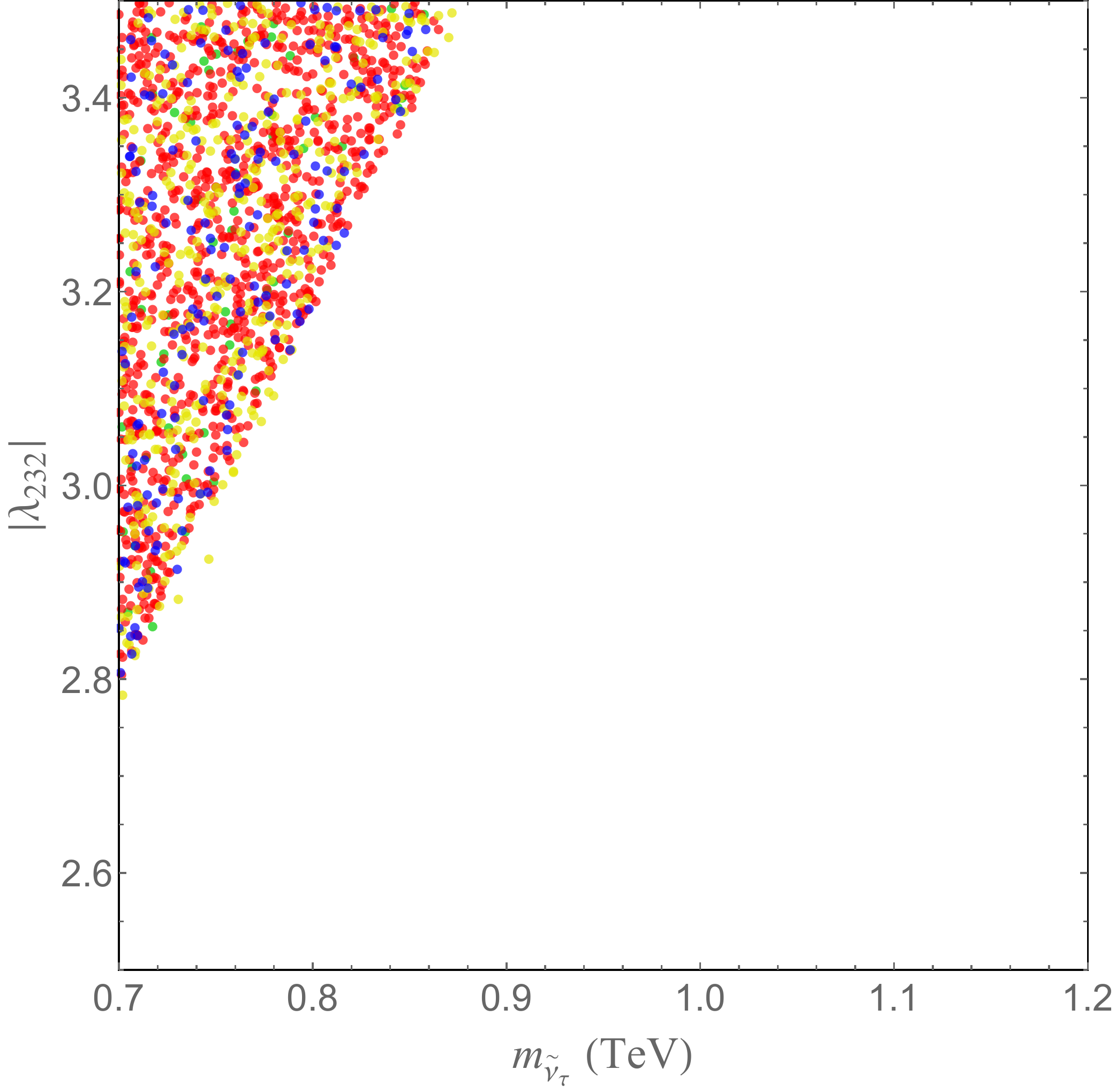}}
    \caption{Scatter plots showing the  correlations between various RPV3 parameters in Eq.~\eqref{eq:BP}. All these points can simultaneously explain $R_{D^{(*)}}$, $R_{K^{(*)}}$ and $(g-2)_\mu$ anomalies at $3\sigma$ CL, while being consistent with all the low-energy and LHC constraints. The yellow (blue) points correspond to $|\lambda^\prime_{232}|<0.2$ and $|\lambda^\prime_{223}|<(>)1$. The  red points correspond to  $|\lambda^\prime_{232}|>0.2$ and $1.5<\lambda^\prime_{223}/\lambda^\prime_{232}<5.5$. The green points correspond to the crossover region from red to blue. The three benchmark points (BP1, BP2, BP3) discussed in the text are taken respectively from the densest regions of the red, yellow and blue solutions.}
    \label{fig:scan}
\end{figure*}

We then perform a random scan over the 6-dimensional parameter space given in  Eq.~\eqref{eq:BP} with the following ranges:
\begin{align}
   & |\lambda_{232}|\in [2.5, \ 3.5], \   
   %\nonumber \\
   % & 
    |\lambda^\prime_{233}|\in [0.01, \ 3], \ \nonumber \\
  &  |\lambda^\prime_{223}|\in [0.01, \ 3], \ 
    |\lambda^\prime_{232}| \in [0.01, \ 3], \  \nonumber \\ 
  & 
 m_{\widetilde{b}_R}\in [1.2, \  10]~{\rm TeV}, \ 
    m_{\widetilde{\nu}_\tau}\in [0.7, \  1.2]~{\rm TeV} , 
\end{align}
and look for solutions that could simultaneously explain $R_{D^{(*)}}$, $R_{K^{(*)}}$ and $(g-2)_\mu$ anomalies at either $2\sigma$ or $3\sigma$, while being consistent with all the low-energy constraints discussed above. Out of 30 million points scanned, we only found 1570 solutions, as shown by the scatter plots in Fig.~\ref{fig:scan}. Note that the lower edges of the $m_{\widetilde{\nu}_\tau}$ and $m_{\widetilde{b}_R}$ scan ranges correspond to the current LHC limits, see Sec.~\ref{sec:collider}.

We divide all the obtained solutions into three characteristically different regions, as shown by the red, blue and yellow points in Fig.~\ref{fig:scan}. In particular, from Fig.~\ref{fig:scan}(c), we see that the yellow (blue) points correspond to $|\lambda^\prime_{232}|<0.2$ and $|\lambda^\prime_{223}|<(>)1$, while the  red points correspond to  $|\lambda^\prime_{232}|>0.2$ and $1.5<\lambda^\prime_{223}/\lambda^\prime_{232}<5.5$, and the green points simply correspond to the crossover region from red to blue. We can immediately make several observations from these plots, as follows: 
\begin{enumerate}
    \item [(i)] From Figs.~\ref{fig:scan}(b,c), we see that there are both yellow and blue solutions with very small (or almost vanishing) $\lambda^\prime_{232}$ which means that  for these points, we can automatically satisfy the $C_9'-C'_{10}$ constraint discussed in main text, as well as the $B_s\to \mu^+\mu^-$ constraint for any value of the stop mass.

    \item [(ii)] From Fig.~\ref{fig:scan}(c), we see that the red points cluster around $\lambda^\prime_{223}/\lambda^\prime_{232} \sim 3$; this helps to avoid the $B_s-\overline{B}_s$ constraint due to an accidental cancellation.   
    
    \item [(iii)] From Figs.~\ref{fig:scan}(a,b,c), we find that $\lambda^\prime_{233}$ and $\lambda^\prime_{223}$ must have  opposite signs. This is mainly needed to make the second term of Eq.~\eqref{eq:C9C10} 
    negative in order to satisfy the $R_{K^{(*)}}$ anomaly. Similarly,  $\lambda^\prime_{223}$ and $\lambda^\prime_{232}$ are preferred to have the same sign to get cancellation in $B_s-\overline{B}_s$ mixing. On the other hand, as Eq.~\eqref{eq:gm2l} 
    suggests and as shown in  Fig.~\ref{fig:scan}(i), the sign of $\lambda_{232}$ does not matter. 
    
    \item [(iv)] According to Fig.~\ref{fig:scan}(i), the different colored points are totally mixed, which implies mutual orthogonality between the $(m_{\widetilde{\nu}_\tau},\lambda)$ and $(m_{\widetilde{b}_R},\lambda^\prime)$ parameter spaces. In other words,  $( m_{\widetilde{\nu}_\tau},\lambda)$ mostly influences the $(g-2)_\mu$ solutions, whereas  $(m_{\widetilde{b}_R},\lambda^\prime)$ influences the  $R_{D^{(*)}}$ and  $R_{K^{(*)}}$ solutions and the low-energy flavor constraints. This is further illustrated in Figs.~\ref{fig:allowed} and \ref{fig:gm}. 
    
    \item [(v)] From Fig.~\ref{fig:scan}(e), we find that $|\lambda^\prime_{223}| \lesssim 0.57 \: (m_{\widetilde{b}_R}/\mathrm{1\ TeV})$, which is mainly due to the $D^0 \rightarrow \mu^+\mu^-$ constraint. Similarly, from Fig.~\ref{fig:scan}(d), we get $|\lambda^\prime_{233}| \lesssim 1.0 \: (m_{\widetilde{b}_R}/\mathrm{1\ TeV})$. These two conditions imply that the $\lambda^\prime$ contributions to $(g-2)_\mu$ in Eq.~\eqref{eq:gm2l} 
    cannot be large; therefore, the bulk of the RPV3 contribution must come from the $\lambda$ sector, which requires relatively larger $\lambda_{232}\gtrsim 2.8$ and smaller $m_{\widetilde{\nu}_\tau}\lesssim 0.9$ TeV (to keep $\lambda_{232}$ perturbative) to satisfy the $(g-2)_\mu$ anomaly, as confirmed in  Fig.~\ref{fig:scan}(i). 
    
    \item [(vi)] From Fig.~\ref{fig:scan}(g), we find that $\sqrt{-\lambda^\prime_{223}\lambda^\prime_{233}} \sim (0.20-0.28)\: (m_{\widetilde{b}_R}/\mathrm{1\ TeV})$, which mainly comes from the  $B \rightarrow K\nu\overline{\nu}$ constraint.

     \item [(vii)] From Fig.~\ref{fig:scan}(a), we find that for the yellow and blue points, $|\lambda^\prime_{233}\lambda^\prime_{223}|$ is small to satisfy the $B_s-\overline{B}_s$ mixing constraint.
     
     \item [(viii)] From Fig.~\ref{fig:scan}(f), we see that  $|\lambda^\prime_{232}| \lesssim 1.5$. Thus, according to Fig.~\ref{fig:scan}(c), $|\lambda^\prime_{232}|$ should be either small (yellow and blue) or $\sim |\lambda^\prime_{223}|/3$ (red). Also, from Fig.~\ref{fig:scan}(d), $|\lambda^\prime_{233}| \gtrsim 0.20$ and from Fig.~\ref{fig:scan}(e), $|\lambda^\prime_{223}| \gtrsim 0.12$. 
     
     \item [(ix)] Figs.~\ref{fig:scan}(d,e,f) suggest that $m_{\widetilde{b}_R}\gtrsim 1.44$ TeV, slightly stronger than the direct LHC bound of 1.23 TeV~\cite{CMS:2017ybq}. 
     
     \item [(x)] Fig.~\ref{fig:scan}(h) gives the range of $X_{bs}X_{\mu\mu}$ for the RPV3 contribution to $R_{K^{(*)}}$, since the second term in Eq.~\eqref{eq:C9C10} 
     gives the correct sign, whereas the first term gives the wrong sign.
\end{enumerate}

\section{Benchmark Points} \label{sec:BP}
We will choose our benchmark points for the collider study in the next section based on the results of our numerical scan in Fig.~\ref{fig:scan} and the  above-mentioned observations. Specifically, we choose three benchmark points (BP1, BP2, BP3), one each from the red (BP1), yellow (BP2) and blue (BP3) solutions found above.     

\begin{figure*}[t!]
    \centering
   \subfigure[~BP1 (Red)]{ \includegraphics[width=0.55\textwidth]{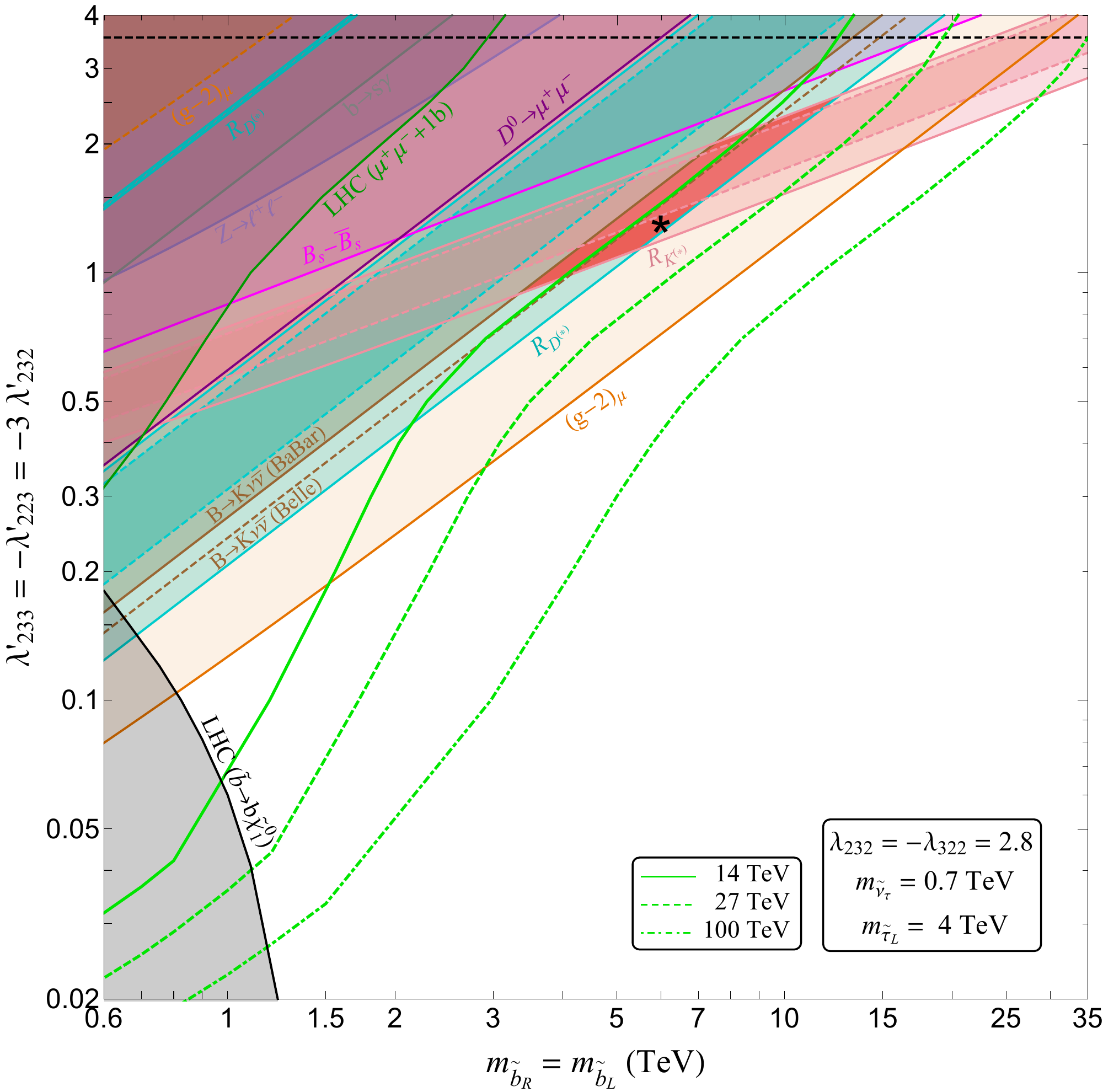}} \\
  \subfigure[~BP2 (Yellow)]{ \includegraphics[width=0.49\textwidth]{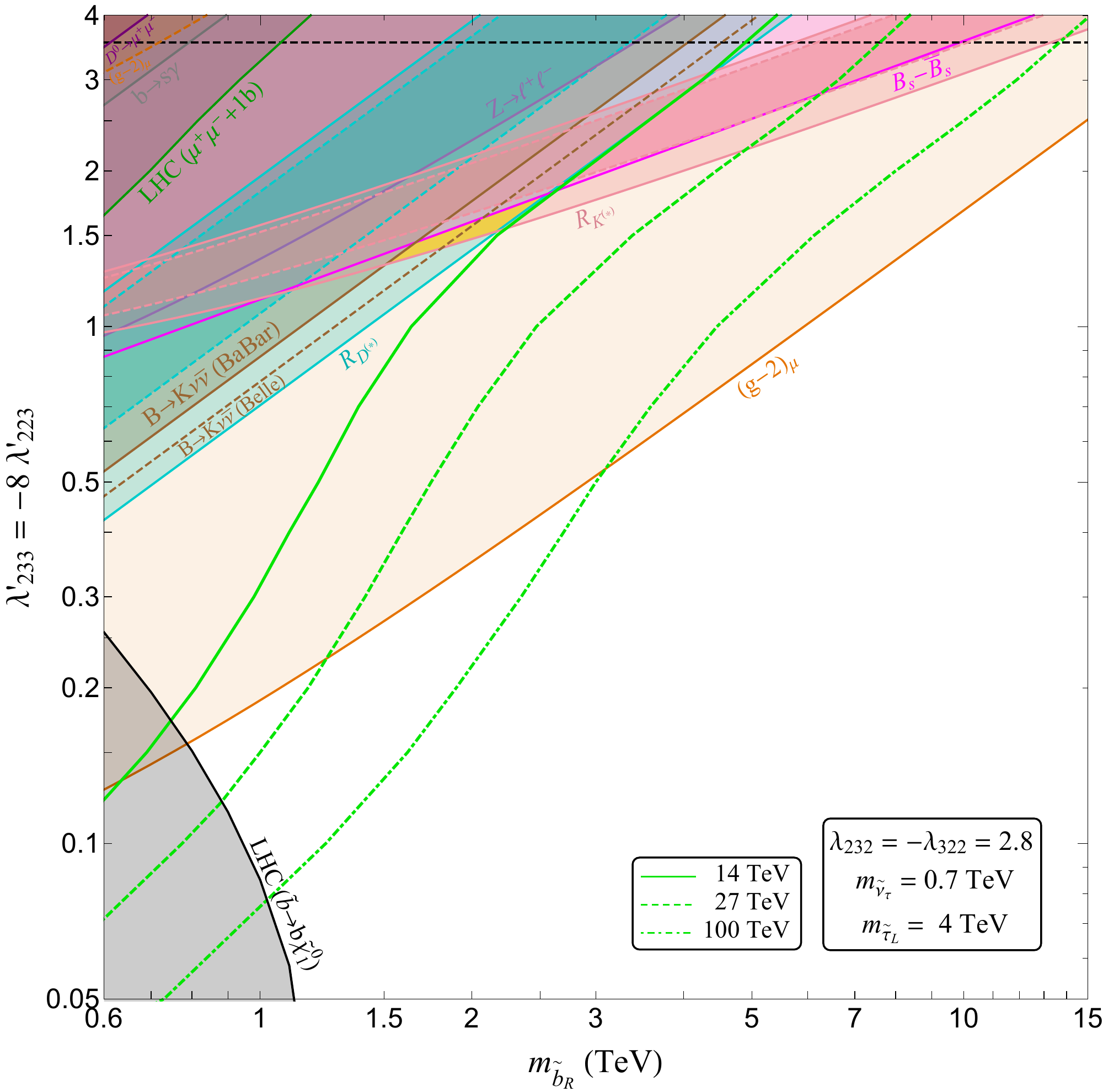}}
   \subfigure[~BP3 (Blue)]{\includegraphics[width=0.49\textwidth]{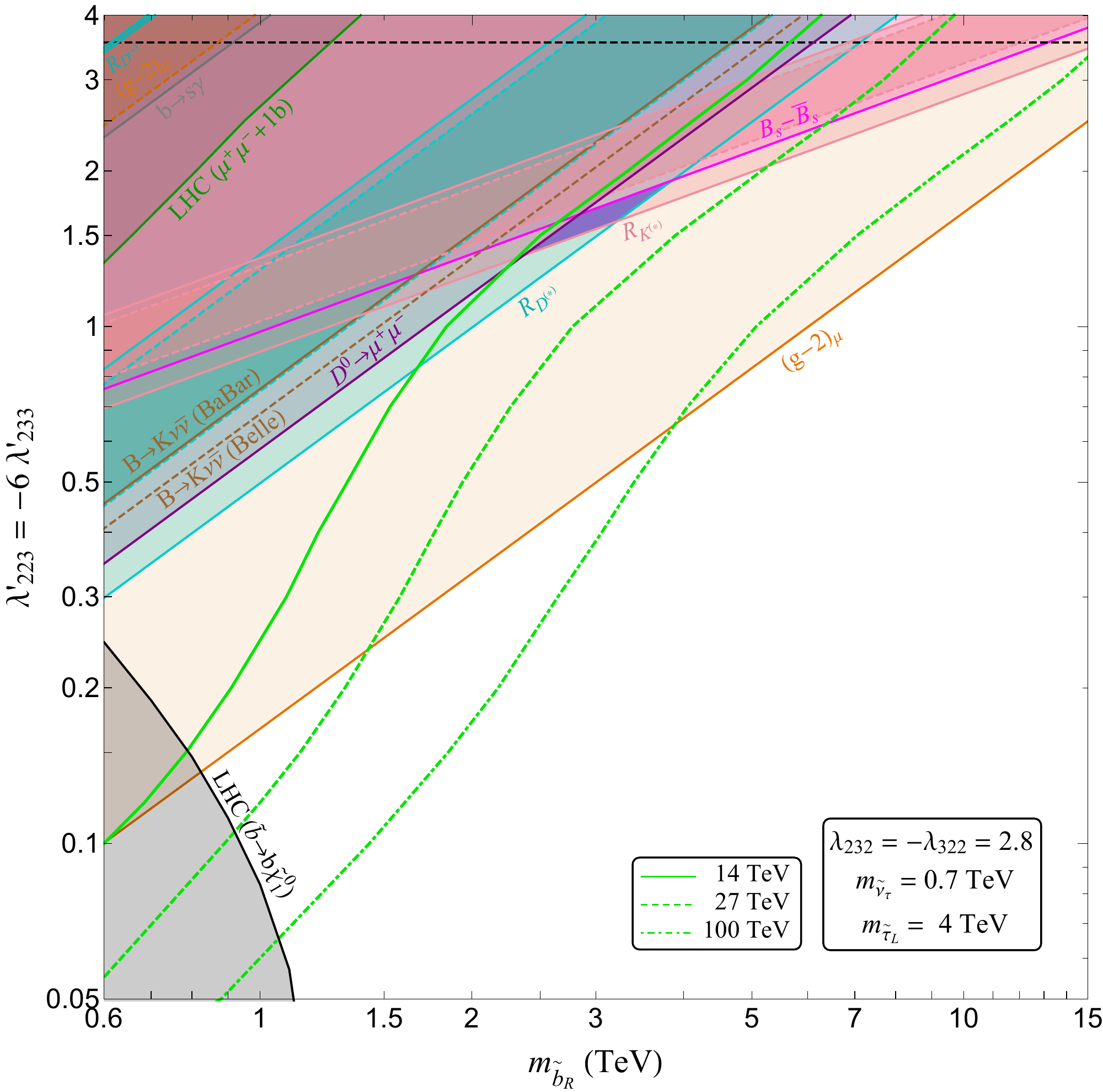}}
    \caption{
    Three RPV3 benchmark cases in the  $(m_{\widetilde b_R}, \lambda^\prime_{233})$ parameter space explaining the flavor anomalies. The cyan, pink and orange shaded regions with solid (dashed) boundaries explain the $R_{D^{(*)}}$, $R_{K^{(*)}}$ and $(g-2)_\mu$ anomalies at $3\sigma$ ($2\sigma$) respectively. The black-shaded region is excluded by the current LHC search for sbottoms in the bottom+neutralino channel, whereas the dark green-shaded region is the LHC exclusion derived from a $\mu^+\mu^-+1b$ search. The horizontal dotted line shows the perturbativity limit of $\sqrt{4\pi}$. 
    Other shaded regions show the relevant low-energy flavor constrains on the parameter space from $B\to K\nu \overline{\nu}$ (brown),  
	$B_s-\overline{B}_s$ mixing (magenta),  
	$D^0\to\mu^+\mu^-$ (purple), $b\to s\gamma$ (grey) and $Z\to \ell^+\ell^-$ (violet). The allowed overlap regions simultaneously explaining the $R_{D^{(*)}}$, $R_{K^{(*)}}$ and $(g-2)_\mu$ anomalies are shown by the red (top), yellow (bottom left) and blue (bottom right) shaded regions for the three benchmark cases. The $*$ mark on the top panel gives representative values of $m_{\widetilde b_R}$ and $\lambda^\prime_{233}$ in the BP1 
	scenario that are used in Fig.~\ref{fig:gm}. The green solid, dashed and dot-dashed contours respectively show the $2\sigma$ sensitivities of the 14 TeV LHC, 27 TeV and 100 TeV $pp$ colliders in the $\overline{t}\mu^+\mu^-$ channel discussed in the text.}
    \label{fig:allowed}
\end{figure*}

\begin{figure}[t!]
    \centering
   \includegraphics[width=0.45\textwidth]{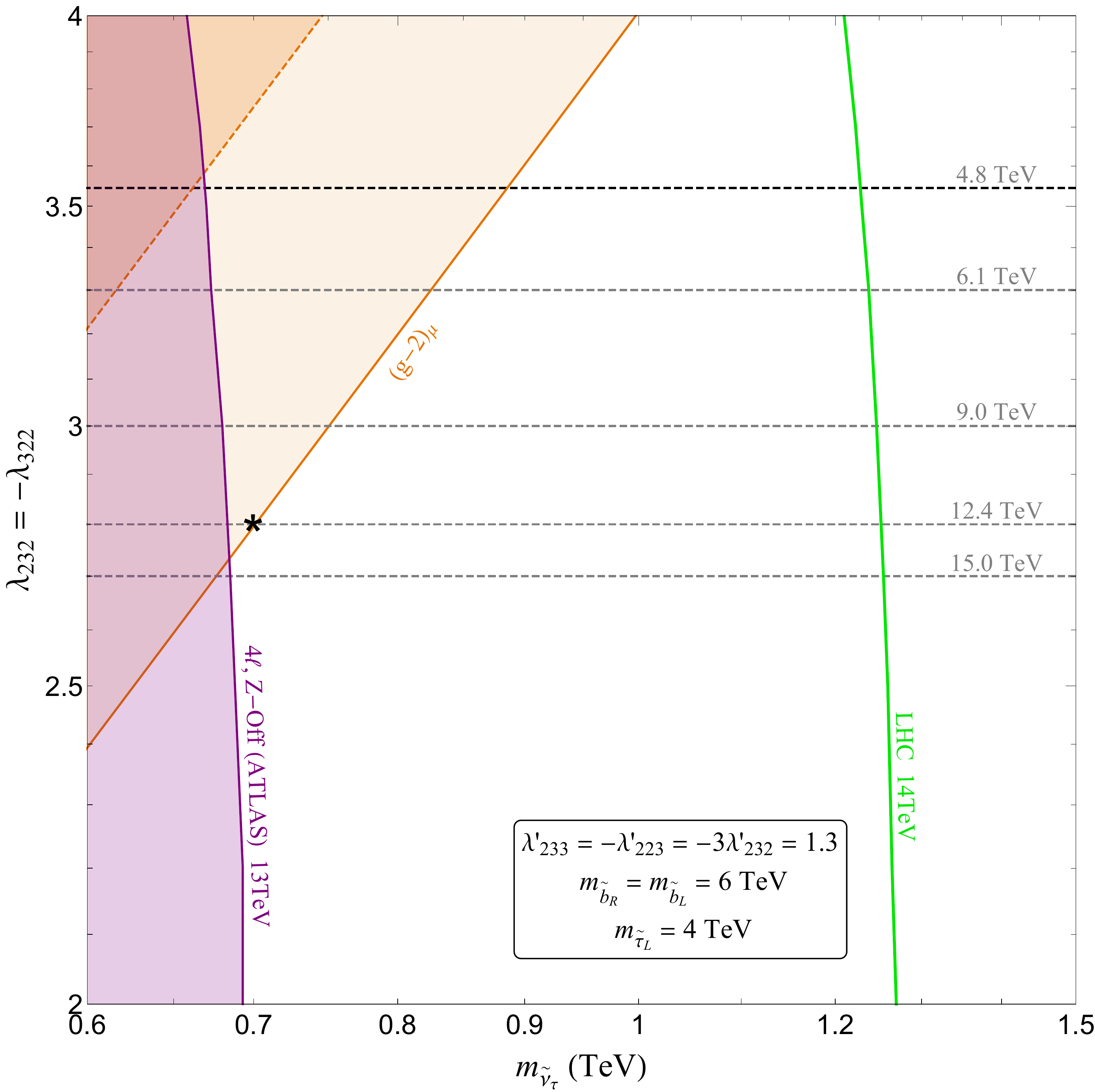}
    \caption{The $(g-2)_\mu$-preferred region (orange-shaded) of the $(m_{\widetilde{\nu}_\tau},\lambda_{232})$ parameter space. The purple-shaded region is excluded by a 13 TeV LHC multi-lepton search~\cite{ATLAS:2021eyc}, whereas the green curve is the 14 TeV HL-LHC sensitivity. The horizontal (gray) dashed lines show the positions of the Landau pole for given $\lambda$-couplings and the black dashed line shows the  perturbativity limit. The $*$ gives representative values of $m_{\widetilde\nu_\tau}$ and $\lambda_{232}$ used in Fig.~\ref{fig:allowed}.}
    \label{fig:gm}
\end{figure}

\begin{itemize}
    \item {\bf BP1 (Red):} $\lambda^\prime_{233} = -\lambda^\prime_{223} = -3\lambda^\prime_{232}$. The allowed region in this case is shown in Fig.~\ref{fig:allowed} (a) 
    by the red shaded region.   
    
    \item {\bf BP2 (Yellow):} $\lambda^\prime_{233} = -8\lambda^\prime_{223}$, $\lambda^\prime_{232} = 0$. The allowed region in this case is shown in Fig.~\ref{fig:allowed} (b) by the yellow shaded region.   
    %From Figure(a), $\lambda^\prime_{233} = -8\lambda^\prime_{223}$ cross the most dense yellow region $\Rightarrow$ more yellow dots $\Rightarrow$ corresponds to a larger yellow region in Figure(f).Once $\frac{\lambda^\prime_{233}}{\lambda^\prime_{223}} \leq -14$, the yellow overlap region in Figure(n) will disappear. Corresponds to a tangent line that touch the yellow region in Figure(a).
    
    \item {\bf BP3 (Blue):} $\lambda^\prime_{223} = -6\lambda^\prime_{233}$, $\lambda^\prime_{232} = 0$. The allowed region in this case is shown in Fig.~\ref{fig:allowed} (c) by the blue shaded region.   
    
\end{itemize} 
The size of the allowed region in each case is directly correlated with the density of the corresponding points in Fig.~\ref{fig:scan}. Therefore, our BP1 is taken from the densest region of the red solution, in order to maximize the overlap region in Fig.~\ref{fig:allowed}. 
For BP2 and BP3, we just choose $\lambda^\prime_{232}=0$ for simplicity. Since the $\lambda$ coupling and the tau-sneutrino mass are relevant only for $(g-2)_\mu$, we fix $\lambda_{232}=-\lambda_{322}=2.8$ and $m_{\widetilde{\nu}_\tau}=0.7$ TeV (see Fig.~\ref{fig:gm}) 
in all three cases  to explain the $(g-2)_\mu$ anomaly at $3\sigma$ ($2\sigma$) CL, as shown by the orange shaded regions with solid (dashed) boundaries in Fig.~\ref{fig:allowed}.

In BP1, since both $\lambda^\prime_{233}$ and $\lambda^\prime_{232}$ are nonzero, there is a lower limit on the stop mass from the $C_9'-C'_{10}$ constraint  [cf.~Eq.~\eqref{eq:primebound}]
:  $m_{\widetilde{t}_L}\gtrsim (14-40)$ 
TeV  for the overlap region.\footnote{According to Ref.~\cite{Buckley:2016kvr}, stop masses lower than about 10 TeV are preferred from naturalness point of view. For a quantitative measure, the level of fine-tuning must be less than some fixed amount, taken there to be the arbitrary threshold of 10\%.} However, this limit does not apply for BP2 and BP3, since $\lambda^\prime_{232}=0$ in those cases; therefore, the stop can be as light as the current LHC bound of $\sim 800$ GeV~\cite{ParticleDataGroup:2020ssz} in these cases. We have also checked that the constraints from $B_s \to \mu^+\mu^-$, whose amplitude is proportional to $C_{10}^\mu-C_{10}^{\prime\mu}$~\cite{Becirevic:2015asa}, is easily satisfied for all three BPs, with the RPV3 contribution to ${\rm BR}(B_s \to \mu^+\mu^-)\lesssim 10^{-12}$, well below the current experimental precision: ${\rm BR}(B_s \to \mu^+\mu^-)_{\rm exp} = \left(2.69^{+0.37}_{-0.35}\right)\times 10^{-9}$ ~\cite{ATLAS:2020acx}. 

We should also comment on the $B_s-\overline{B}_s$ mixing constraint. For our benchmark points, the last term in Eq.~\eqref{eq:bbmixing} does not contribute, as both $\lambda^\prime_{332}$ and $\lambda^\prime_{323}$ are set to zero.\footnote{As discussed before, a non-zero $\lambda^\prime_{3ij}$ combined with $\lambda_{232}$ and light tau-sneutrino propagator will lead to strong tree-level meson decays $\overline{d}_i d_j \to \mu^+ \mu^-$. The excellent agreement between the experimental measurement~\cite{ATLAS:2020acx} and SM prediction~\cite{Bobeth:2013uxa} of ${\rm BR}(B_s^0 \to \mu^+ \mu^-)$ requires an almost zero last term in Eq.~\eqref{eq:bbmixing}.} For BP1 with $\lambda^\prime_{223}/\lambda^\prime_{232}=3\simeq -P_1^{LR}/P_1^{VLL}$, there is a cancellation (at the level of 5\%) between the other two terms in Eq.~\eqref{eq:bbmixing}, thus enabling us to explain $R_{K^{(*)}}$ within $1\sigma$, while this is not the case in BP2 and BP3 where $R_{K^{(*)}}$ can only be explained at $3\sigma$ level. Note that one can always assume a non-zero $\lambda^\prime_{232}$ for BP2 and BP3 (corresponding to the yellow and blue points out of the vertical axis in Figs.~\ref{fig:scan}(b,c)) to make $B_s-\overline{B}_s$ mixing constraint weaker and enlarge the allowed parameter space for $R_{K^{(*)}}$, but this makes the $B \to K\nu\overline{\nu}$ constraint stronger, which limits the allowed region in BP2.

%\clearpage
Our fit results for the best-case scenario are shown in Figs.~\ref{fig:allowed} and \ref{fig:gm} for the mutually orthogonal parameter spaces of $(m_{\widetilde{b}_R},\lambda^\prime)$ and  $(m_{\widetilde{\nu}_\tau},\lambda)$ respectively. In Fig.~\ref{fig:allowed}, the cyan, pink, and orange-shaded regions with solid (dashed) boundaries explain the $R_{D^{(*)}}$, $R_{K^{(*)}}$, and $(g-2)_\mu$ anomalies respectively at $3\sigma$ ($2\sigma$) CL. 
The black-shaded region is the 13 TeV LHC exclusion derived from a sbottom search in the bottom+neutralino channel~\cite{CMS:2017ybq}. The dark-green-shaded region is the 13 TeV LHC exclusion derived from a $\mu^+\mu^-+1b$ search~\cite{ATLAS:2021mla} that is also applicable to our RPV3 scenario; see Sec.~\ref{sec:collider}. The horizontal dashed line shows the perturbativity limit of $\sqrt{4\pi}$. Other shaded regions in Fig.~\ref{fig:allowed} show the relevant low-energy flavor constrains on the $(m_{\widetilde{b}_R},\lambda^\prime_{233})$ parameter space from $B\to K\nu \overline{\nu}$ (brown),  
	$B_s-\overline{B}_s$ mixing (magenta),  
	$D^0\to\mu^+\mu^-$ (purple), $b\to s\gamma$ (grey) and $Z\to \ell^+\ell^-$ (violet); see Appendix~\ref{app:constraints} for more details. 
	
Now turning to the $(m_{\widetilde{\nu}_\tau},\lambda_{232})$ parameter space, the  $(g-2)_\mu$-preferred region at $3\sigma$ ($2\sigma$) is shown by the solid (dashed) orange contours in Fig.~\ref{fig:gm}. We have fixed the other RPV3  parameters using a benchmark point from the allowed region in Fig.~\ref{fig:allowed} as shown by the $*$ mark. The purple-shaded region is excluded by recasting the results of a recent 13 TeV LHC multi-lepton search~\cite{ATLAS:2021eyc}, whereas the green curve is the 14 TeV HL-LHC sensitivity; see Sec.~\ref{sec:collider}. The horizontal black dashed line shows the perturbativity limit of $\sqrt{4\pi}$ as before. Because of the orthogonality between the  $(m_{\widetilde{b}_R},\lambda^\prime)$ and $(m_{\widetilde{\nu}_{\tau}},\lambda)$ parameter spaces, the position of the $*$ in Fig.~\ref{fig:gm} will not change much for BP2 and BP3; therefore, we do not include the corresponding figures for BP2 and BP3.

Since the required $\lambda$-couplings are fairly large in our scenario, we also show the Landau pole positions by the horizontal gray dashed lines, which are obtained by numerically solving the relevant one-loop RG equations (RGEs)~\cite{Allanach:1999mh}. Because the non-zero $\lambda^\prime$ couplings in our scenario do not couple to the third-generation slepton or sneutrino, the RGE for the $\lambda_{232}$ coupling (and similarly, for the $\lambda_{322}$ coupling) is very simple at one-loop level:
\begin{align}
\frac{\rm d}{{\rm d} t} \lambda_{232} \simeq \frac{\lambda_{232}}{16\pi^2}\left( 4\lambda_{232}^2  - \frac{9}{5}g_1^2 - 3g_2^2 \right) \approx \frac{1}{4\pi^2}\lambda_{232}^3 \,,
\end{align}
where $g_1$ and $g_2$ are the $U(1)_Y$ and $SU(2)_L$ gauge couplings, respectively, both of which are much smaller than the $\lambda_{232}$ coupling in the parameter space shown in Fig.~\ref{fig:gm}. We find that the benchmark point shown in Fig.~\ref{fig:gm} hits the Landau pole at 12.4 TeV. 

%\begin{figure}[t!]
%		\centering
%		\includegraphics[width=0.45\textwidth]{final_plots/constraints_gminus2_another_lambda233.pdf}
%	\caption{The $(g-2)_\mu$-preferred region (orange-shaded) of the $(m_{\widetilde{\nu}_\tau},\lambda_{233})$  parameter space. The purple-shaded region is excluded by a 8 TeV LHC multi-lepton search~\cite{ATLAS:2014pjz}.}
%	\label{fig:gm2}
%\end{figure}

 One possible way out of the Landau pole issue is to consider a non-zero $\lambda_{233}$ coupling instead of $\lambda_{232}$. This does not affect the $R_{D^{(*)}}$ and $R_{K^{(*)}}$ fit results presented in Fig.~\ref{fig:allowed} because of the orthogonality between the $\lambda'$ and $\lambda$ parameter space mentioned before. As for the $(g-2)_\mu$-preferred region in the $(m_{\widetilde{\nu}_\tau},\lambda_{233})$ parameter space, the main difference with respect to Fig.~\ref{fig:gm} is that the allowed $\lambda$ couplings can now be as low as 0.8, thus pushing the Landau pole to as high as $2.5\times 10^{16}$ GeV. The reason is that the relevant LHC constraint for $\lambda_{233}$ comes from $\mu^+\mu^-\tau^+\tau^-$ final state (in contrast with the  $\mu^+\mu^-\mu^+\mu^-$ final state in Fig.~\ref{fig:gm}). We did not find any 13 TeV LHC analysis in this channel, and using the old 8 TeV analysis from Ref.~\cite{ATLAS:2014pjz}, we obtain a lower bound of only 150 GeV on the sneutrino mass. A dedicated 13 TeV analysis to update this bound is currently underway.

\section{Collider Signals}\label{sec:collider}
Simple crossing symmetry arguments have been used to establish high-$p_T$ model-independent tests of the $R_{D^{(*)}}$ and $R_{K^{(*)}}$ anomalies in the CMS and ATLAS experiments~\cite{Altmannshofer:2017poe, Faroughy:2016osc, Greljo:2017vvb, Afik:2018nlr, Greljo:2018tzh, Afik:2019htr, Altmannshofer:2020axr}. The basic idea is that the underlying quark-level processes $b\to c\tau \nu$ for $R_{D^{(*)}}$ and $b\to s\ell^+\ell^-$ for $R_{K^{(*)}}$ necessarily imply, by crossing symmetry, the existence of processes like $pp\to \tau \nu$, $pp\to \ell^+\ell^-$, $pp\to b\tau \nu$ and $pp\to b\ell^+\ell^-$, which can be searched for in the high-$p_T$ LHC experiments. In fact, a recent CMS study has found a mild discrepancy in the ratio of differential $\mu^+ \mu^-$ to $e^+ e^-$ pair-production cross sections~\cite{CMS:2021zil}, which might turn out to be important for the $R_{K^{(*)}}$ anomaly. However, the model-independent effective field theory  treatments relating the low-energy operators to the high-$p_T$ LHC signals might break down, if the new physics cut-off scale is smaller than the LHC energies. Thus, it is important to explore all possible high-$p_T$ LHC signals in the context of a given BSM scenario in order to distinguish it from other BSM interpretations of  the flavor anomalies.  

\begin{figure*}[t!]
    \centering
    \subfigure[]{\includegraphics[width=0.3\textwidth]{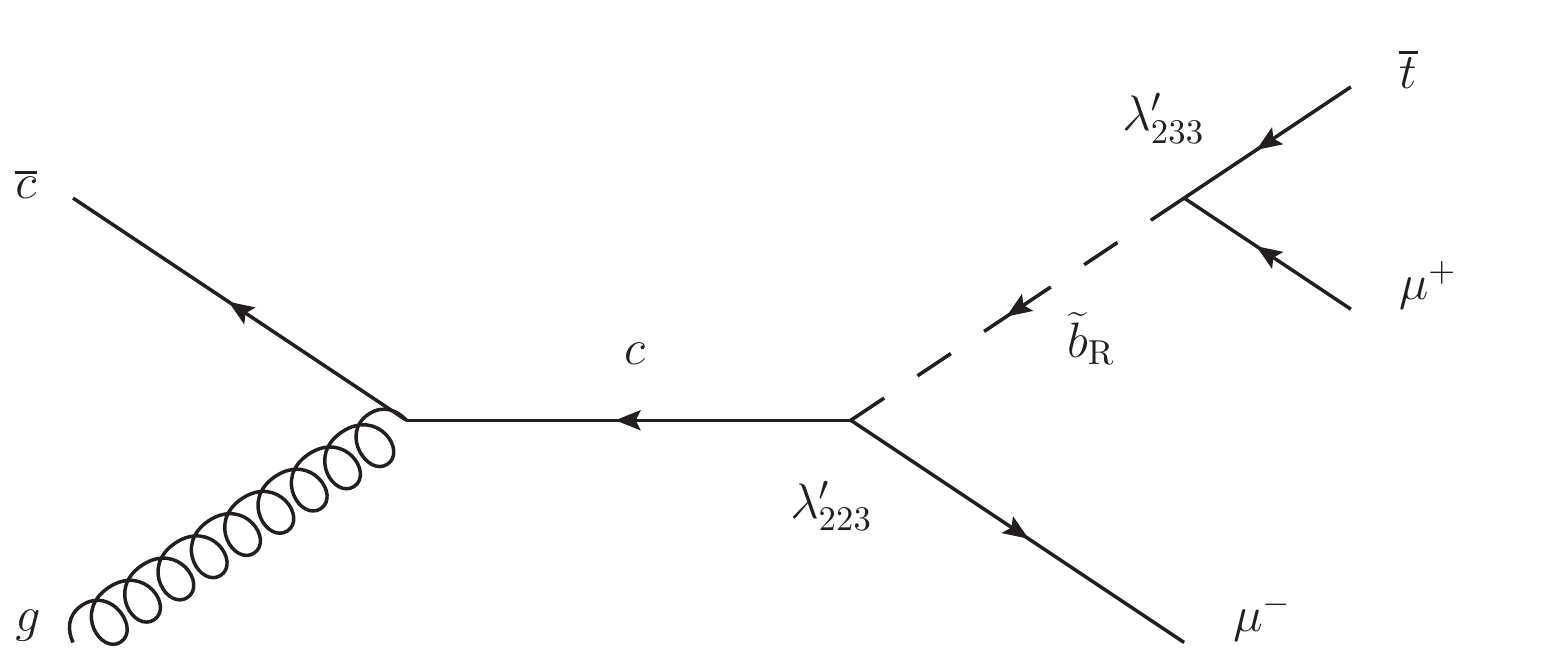}}
    \quad
    \subfigure[]{\includegraphics[width=0.3\textwidth]{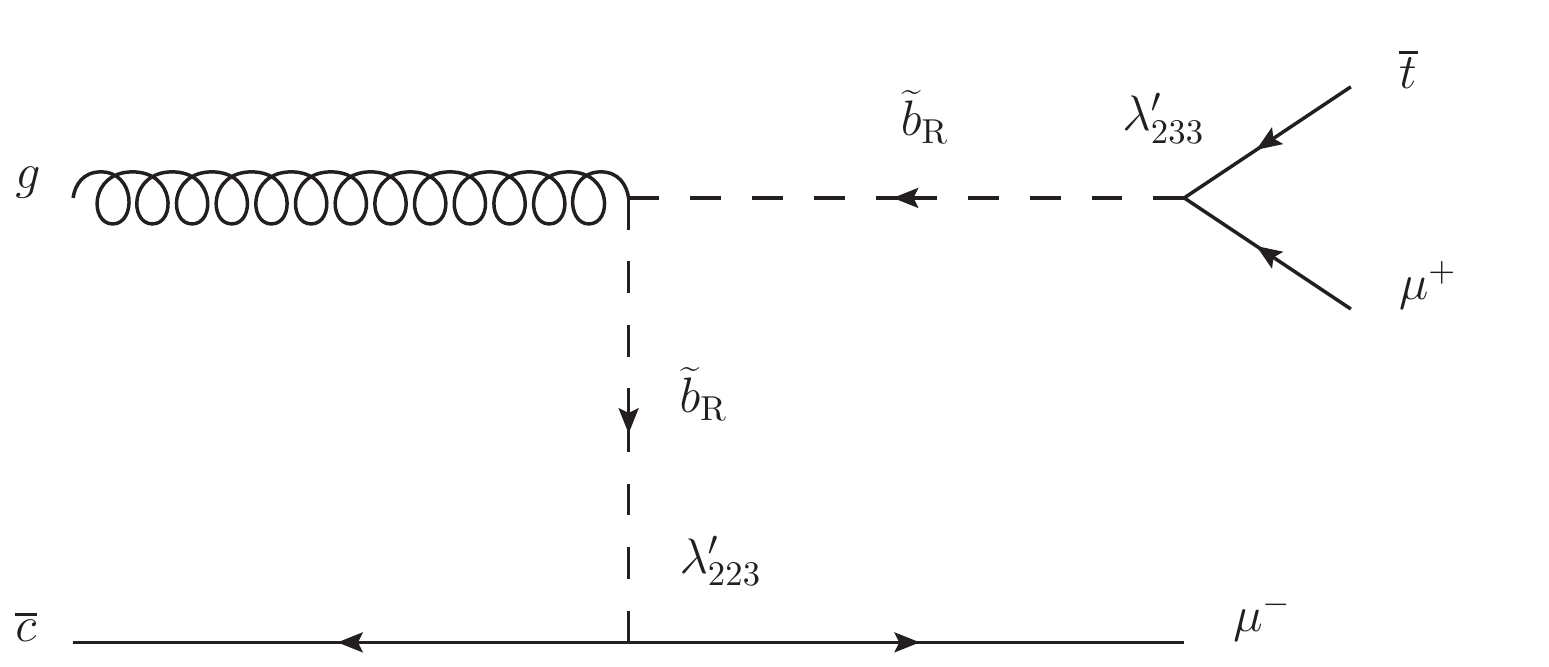}}
    \quad
    \subfigure[]{\includegraphics[width=0.3\textwidth]{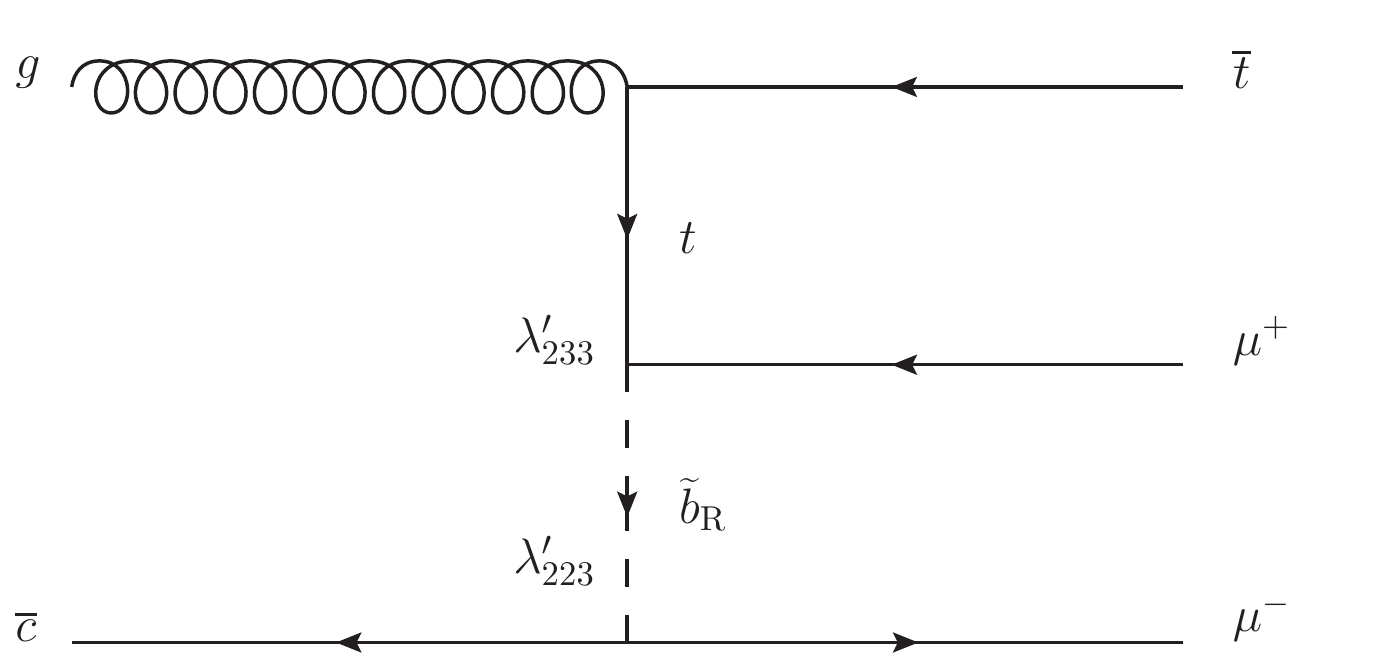}}
    \caption{Representative Feynman diagrams for the signal process $pp\rightarrow\overline{t}\mu^{+}\mu^{-}$. There are similar diagrams for the process $pp\rightarrow t\mu^{+}\mu^{-}$, however the SM background is larger for top-quark final states, compared to the anti-top, so we only consider the latter case for drawing the sensitivity contours in Fig.~\ref{fig:allowed}.}
    \label{fig:signal1}
\end{figure*}

To this effect, we propose some striking LHC signals that could be used as an independent probe of the allowed RPV3 parameter space shown in Figs.~\ref{fig:allowed} and \ref{fig:gm} at the high-$p_T$ LHC and future colliders. For the $(m_{\widetilde{b}_R},\lambda^\prime)$ parameter space in Fig.~\ref{fig:allowed}, we propose the process $pp\to \overline{t}(t)\mu^+\mu^-$ mediated by an sbottom;  see Fig.~\ref{fig:signal1}. 
There is no $pp\to \overline{t}(t)\mu^+\mu^-$ final state in the SM, so the dominant SM background comes from $pp\to \overline{t}(t)\mu^+\mu^-X$ where $X$ can be either a light jet ($j$), $b$-jet, or a gauge boson decaying to jets or leptons, which are somehow missed in the detector. We perform a  parton-level simulation for the signal and background processes using \textsc{MadGraph5\_aMC@NLO v2.7.0}\xspace~\cite{Alwall:2014hca}. We assume that the (anti)top quark can be identified from its decay products and use the minimal trigger cuts  $p_T^{t,\mu}>20$ GeV, $|\eta^{t,\mu}|<2.5$, $\Delta R^{\mu\mu}>0.4$ and $\Delta R^{t\mu}>0.4$ for the $t(\overline{ t})\mu^+\mu^-X$ final states. To estimate the SM background, we further require that the $X$ containing jets, leptons or missing transverse energy is soft enough to evade detection, i.e.~$p_T^{j,b,\ell}<20$ GeV and $\slashed{E}_T<20$ GeV.  

For the RPV3 signal, we use the three benchmark points discussed in the previous section. It is easy to see that the $pp \rightarrow t \mu^{+} \mu^{-}$ and $pp \rightarrow \overline{t} \mu^{+} \mu^{-}$ give similar cross-sections for the signal, because in all three cases, $\lambda^{\prime}_{k13}=0$, and therefore, the first-generation quarks do not participate in the initial state. For our parameter choice, the main contribution comes from $\overline{c}(c) g$ initial states as shown in Fig.~\ref{fig:signal1}. Since the $\overline{c}$ and $c$ contents in proton are very similar, the resulting cross-section should also be the same. However, for the SM background, $t \mu^{+} \mu^{-}X$ final state has two times more background than $\overline{t} \mu^{+} \mu^{-}X$, which mainly comes from the fact that the $u$ content in proton is much larger than the $\overline{u}$ content. Therefore, we will only consider the $pp \rightarrow \overline{t} \mu^{+} \mu^{-}$ final state to show our sensitivity contours. 

\begin{figure}[t!]
    \centering
   \includegraphics[width=0.49\textwidth]{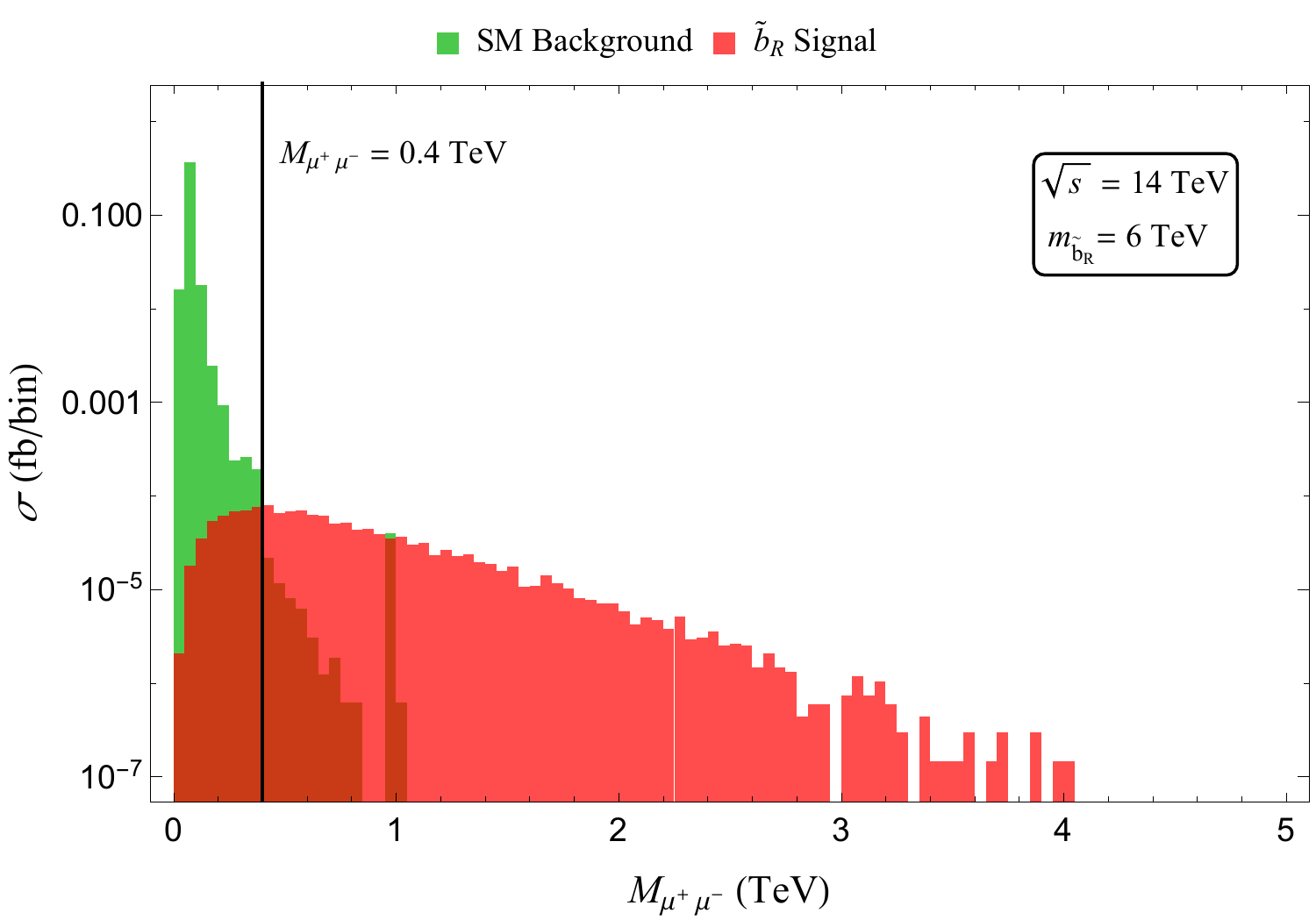}
    \caption{Dimuon invariant mass distribution for the RPV3 signal at the $*$ point in BP1 (red) and SM background (green) in the $pp\to \overline{ t} \mu^+\mu^-$ channel at 14 TeV LHC. }
    \label{fig:dist}
\end{figure}

With the basic trigger cuts, we find that the total SM background for the $pp \rightarrow \overline{t} \mu^{+} \mu^{-}(+X)$ final state at $\sqrt s=14$ TeV is 0.4 fb, which is dominated by $X=j$. For comparison, the corresponding signal cross section for the $*$ point in BP1 in Fig.~\ref{fig:allowed} 
is only $1.5\times 10^{-3}$ fb. However, we can improve the signal-to-background substantially by using their different kinematic features. First of all, the $\mu^+\mu^-$ in the SM background case mainly comes from $Z$ decay, so we expect the dimuon invariant mass $M_{\mu^+\mu^-}$ to peak at the $Z$-mass and to drop significantly at higher masses; see Fig.~\ref{fig:dist}  (green). On the other hand, in our RPV3 case, one of the muons in the final state comes from sbottom decay, so we expect a longer tail in the $M_{\mu^+\mu^-}$ distribution, as confirmed in Fig.~\ref{fig:dist}  (red). Therefore, using an appropriate cut on $M_{\mu^+\mu^-}>400$ GeV, we can maximize the signal-to-background ratio. We find that the corresponding signal at the $*$ point in BP1 and background cross sections after the $M_{\mu^+\mu^-}$ cut are respectively $1.1\times 10^{-3}$ fb and $4.2\times 10^{-4}$ fb. Further improvements in the signal-to-background can in principle be achieved using the fact that for an on-shell sbottom decaying to $\overline{t}\mu^+$, we expect a peak at the sbottom mass in the invariant mass $M_{\overline{t}\mu^+}$  distribution for the signal, but not for the background. However, since the final-state reconstruction involving top quarks is somewhat involved, especially for the leptonic decay of the $W$ boson coming from the top, and also the sbottom mass is not known a priori (we use it as a free parameter in Fig.~\ref{fig:allowed}), 
we refrain from using the $M_{\overline{t}\mu^+}$ cut in our analysis.

Assuming an integrated luminosity of ${\cal L}=3000~{\rm fb}^{-1}$, we show the $2\sigma$ signal significance in Fig.~\ref{fig:allowed} by the green solid, dashed and dot-dashed contours for $\sqrt s=14$, 27 and 100 TeV colliders, respectively. We find that a portion of the overlap region explaining all flavor anomalies can already be accessed at the HL-LHC, while the proposed future colliders should be able to access the entire allowed parameter space.  

\begin{figure}[t!]
    \centering
   \includegraphics[width=0.3\textwidth]{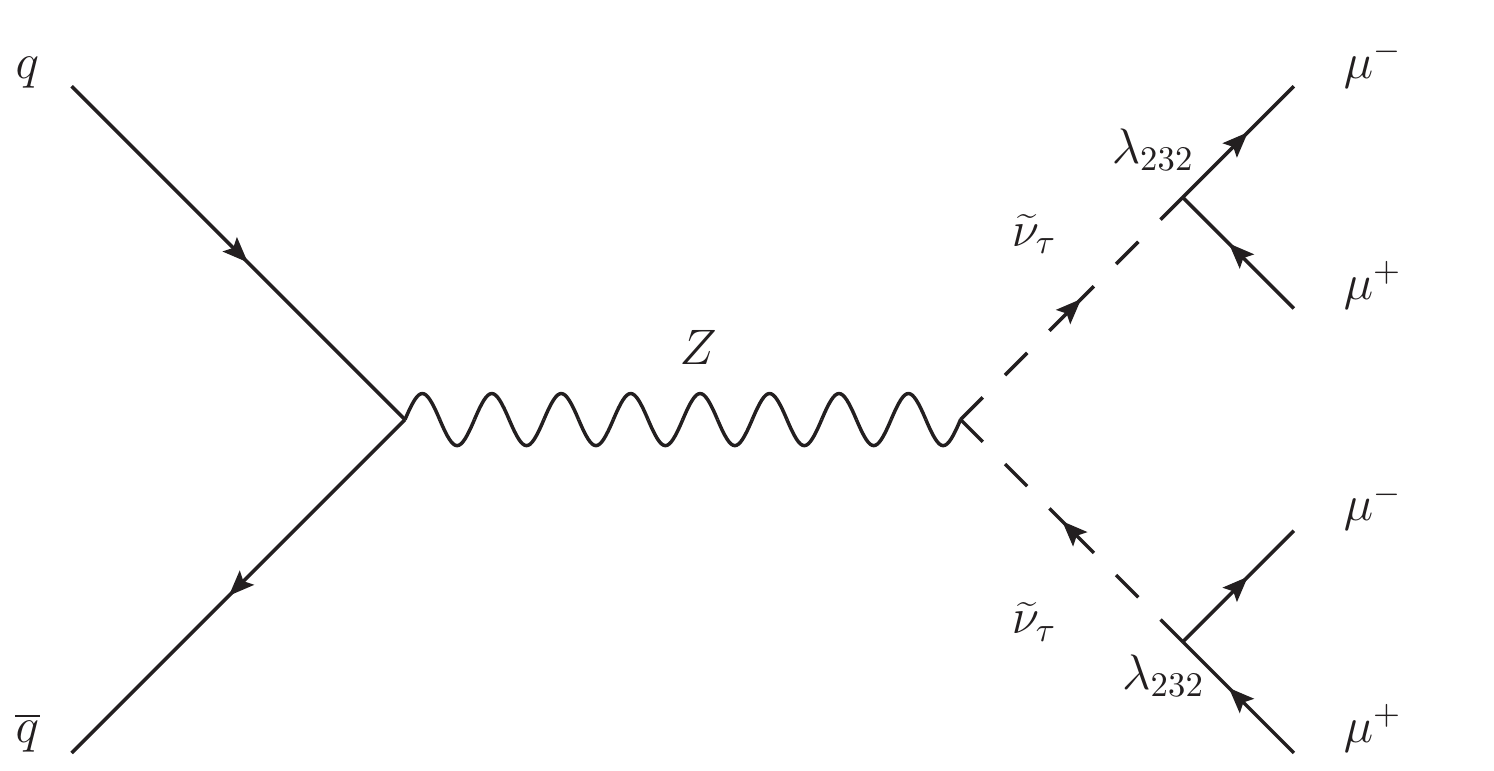}
    \caption{Feynman diagram for the four-lepton signal from the sneutrino pair-production in our RPV3 model. }
    \label{fig:signal2}
\end{figure}

Now for the $(m_{\widetilde{\nu}_\tau},\lambda)$ parameter space in Fig.~\ref{fig:gm} relevant only for the $(g-2)_\mu$ anomaly, we focus on the spectacular four-muon final state~\cite{Chakraborty:2015bsk} coming from the sneutrino pair-production, followed by each sneutrino decaying into two muons via the $\lambda_{232}$ coupling; see Fig.~\ref{fig:signal2}. Such multilepton channels are very clean even at the hadron colliders, and in fact, the results of a recent ATLAS multilepton analysis~\cite{ATLAS:2021eyc} can already be recast into a new bound on the sneutrino mass. Using the 95\% CL observed cross section limit of $0.044$ fb for the $4\ell$, off-$Z$ signal region with $M_{4\mu}>400$ GeV\footnote{This also removes potential contributions from heavy neutral Higgs to $ZZ$ final states.} and the same selection criteria as in Ref.~\cite{ATLAS:2021eyc}, 
%i.e.,~$p_T^\mu>25$ GeV and $|\eta^\mu|<2.47$, 
we obtain a lower bound of $m_{\widetilde{\nu}_\tau}\gtrsim 670$ GeV, as shown by the purple-shaded region in Fig.~\ref{fig:gm}. 
%Note that this is independent of the $\lambda$ coupling in the region shown here, because for such large values of $\lambda_{232}$, the BR of $\widetilde{\nu}_\tau\to \mu^+\mu^-$ is more than 95\%. 
This LHC constraint already rules out a big chunk of the $(g-2)_\mu$-preferred region and pushes the $\lambda_{232}$ coupling toward the perturbativity limit. The HL-LHC can completely cover the remaining $(g-2)_\mu$-preferred region, as shown by the green curve in Fig.~\ref{fig:gm}.

\section{Discussion}\label{sec:dis}
In this section, we make a few remarks on our results before concluding our discussion. 
\subsection{Interplay between Anomalies}
We find in Fig.~\ref{fig:gm} 
that only a narrow region in the $(m_{\widetilde \nu}, \lambda_{232})$ parameter space is allowed that could explain the $(g-2)_\mu$ anomaly in our minimal RPV3 setup. It is worth checking how does the $(g-2)_\mu$-preferred region gets affected on its own, and more importantly, if dropping the $R_{D^{(*)}}$ and  $R_{K^{(*)}}$ anomalies could open up more parameter space in Fig.~2~\ref{fig:gm}. 
To this effect, we find that allowing $\lambda^\prime_{213} \neq 0$ in Eq.~\eqref{eq:gm2l}, 
the $(g-2)_\mu$ solution can be  improved only slightly, compared to Fig.~\ref{fig:gm}. 
With $m_{\widetilde{b}_R} = 1.23$ TeV (the minimum value allowed by LHC 13 TeV data) and $|\lambda^\prime_{233}| = 1.57$ (the maximum value allowed from the $Z$-decay universality constraint $g^\mu_A/g^e_A$, where $g^\ell_A$ is the axial-vector coupling of $Z$ to leptons), the lower $3\sigma$ bound of $(g-2)_\mu$ starts at $(0.7 \mathrm{~TeV},2.66)$ and ends at $(0.93 \mathrm{~TeV},\sqrt{4\pi})$ in the $(m_{\widetilde{\nu}_\tau},\lambda_{232})$ parameter space. The lower $2\sigma$ bound is still not reached for $\lambda_{232} < \sqrt{4\pi}$. Because of the constraints $B \to K\nu\overline{\nu}$ and $K \to \pi\nu\overline{\nu}$, $\lambda^\prime_{213},\lambda^\prime_{223} \approx 0$ and thus cannot contribute much to the $(g-2)_\mu$ anomaly in this optimal $(g-2)_\mu$ case.

Nevertheless, it is important to mention that due to the orthogonality between $(m_{\widetilde{\nu}_\tau},\lambda)$ and $(m_{\widetilde{b}_R},\lambda^{\prime})$ parameter spaces in our RPV3 scenario, even if the four-muon signal completely rules out the $(g-2)_\mu$-favored region in Fig.~\ref{fig:gm} 
(drawn for BP1, but similar for BP2 and BP3 as we can see from Fig.~\ref{fig:scan}(i)), the $R_{D^{(*)}}$ and $R_{K^{(*)}}$ anomalies can still be explained by the $(m_{\widetilde{b}_R},\lambda^{\prime})$ parameter space shown in Fig.~\ref{fig:allowed}. Similarly, suppose the $pp\to \overline{t}\mu^+\mu^-$ signal completely rules out one of the overlap regions in  Fig.~\ref{fig:allowed}, but it will not affect the $(g-2)_\mu$ solution in Fig.~\ref{fig:gm}.  

\subsection{Leptoquark versus RPV3} 
Several BSM scenarios have been invoked to explain the flavor anomalies, but very few have the ability to explain all the flavor anomalies simultaneously in a minimal, theoretically well-motivated setup like the RPV3. Leptoquarks (LQs) have been a popular choice, but a single scalar LQ solution has now been disfavored by global fits~\cite{Angelescu:2021lln}. A single vector LQ $U_1({\bf 3}, {\bf 1}, 2/3)$ still remains a viable option~\cite{Calibbi:2015kma, Bhaskar:2021pml, Ban:2021tos}, but must be embedded in some ultraviolet completion like the Pati-Salam gauge group~\cite{Calibbi:2017qbu, Barbieri:2017tuq, Blanke:2018sro}, thus necessarily requiring more particles to cancel gauge anomalies, and hence, losing its minimality feature. Another alternative is to invoke more than one scalar LQs~\cite{Chen:2017hir, Crivellin:2017zlb, Bigaran:2019bqv, Crivellin:2019dwb, Saad:2020ihm, Babu:2020hun}. The right-sbottom  with $\lambda^\prime$ couplings in our RPV3 scenario behaves exactly like the $SU(2)_L$-singlet LQ $S_1({\bf 3}, {\bf 1}, -1/3)$ originally invoked in Ref.~\cite{Bauer:2015knc}, which still gives an excellent fit to the $b\to c\tau\nu$ data, including polarization observables and forward-backward asymmetry~\cite{Carvunis:2021dss}. However, the same $S_1$ LQ cannot explain the $b\to s\mu^+\mu^-$ data simultaneously~\cite{Angelescu:2018tyl}, while being consistent with the low-energy constraints, in particular from $B_s-\overline{B}_s$ mixing. This is a key difference with RPV3, where a TeV-scale sbottom by itself can explain both $R_{D^{(*)}}$ and $R_{K^{(*)}}$, owing to a (partial) cancellation in the $B_s-\overline{B}_s$ mixing [cf.~Eq.~\eqref{eq:bbmixing}]. 
Another important difference is the $\lambda$ coupling, which gives rise to the distinct four-lepton signal in the RPV3 scenario and uniquely distinguishes our scenario from LQ models.

\subsection{Precision  Tests} 
Apart from the collider tests proposed here, our RPV3 solution to the flavor anomalies can also be probed via low-energy precision observables at LHCb and Belle-II. For instance, the $*$ benchmark point in Fig.~\ref{fig:allowed} predicts the ratio [cf.~Eq.~\eqref{eq:RBKvv}] $R_{B\to K\nu\overline{\nu}}=2.1$, which is just below the Belle 95\% CL upper limit of  3.2~\cite{Belle:2017oht,Buras:2014fpa}. The future Belle-II sensitivity can improve this limit by up to a factor of 5~\cite{Belle-II:2018jsg}, which should be able to completely probe the overlap region. In particular, the red, yellow, and blue overlap regions in Fig.~\ref{fig:allowed} can be completely excluded for $R_{B\to K\nu\overline{\nu}}<1.7$, 2.0 and 1.1 respectively. This is a distinct feature of our RPV3 scenario.\footnote{For instance, in the $U_1$ vector LQ case, there is no tree-level contribution to $B\to K\nu\overline{\nu}$ and any prediction involving loop-processes depends on the UV-completion details.} Similarly, future lattice improvements in the precision of the SM prediction for $B_s-\overline{B}_s$ mixing could be fateful for the overlap region in Fig.~\ref{fig:allowed}. In addition, hints of LFUV in other independent observables involving the third generation, such as LFV $\tau$ and $B$ decays, $b\to c\mu\nu/b\to ce\nu$ and  baryonic decay modes like $\Lambda_b\to \Lambda \ell^+\ell^-$, would provide critical further tests of our proposal. Furthermore, there are other lepton flavor violating $B$ and $\tau$ decays that could get enhanced contributions from RPV3 within reach of Belle II sensitivity~\cite{Altmannshofer:2020axr}. If the flavor anomalies persist and grow in statistical significance, the precision flavor observables mentioned above, in conjunction with the collider observables discussed in Sec.~\ref{sec:collider}, might be able to uniquely distinguish our RPV3 interpretation from other BSM interpretations. 

\subsection{Caveats}
In spite of all the above-mentioned attractive features of our RPV3 scenario, there are a few weak points which we just lay out here for future contemplation. 

\noindent
$\bullet$ {\bf Landau Pole:} In the minimal RPV3 setup presented here, some of the $\lambda'$ and $\lambda$ couplings are required to be fairly large $\gtrsim {\cal O}(1)$. Such large couplings would hit the Landau pole very quickly, preventing the model from being valid all the way up to the gauge coupling unification scale. For instance, the benchmark point shown in Fig.~\ref{fig:gm} hits the Landau pole at 12.4 TeV. There might be a way out in the general RPV-MSSM with more parameters, but a detailed analysis of the full MSSM parameter space is beyond the scope of this work. \\
 \noindent   
$\bullet$ {\bf Neutrino Mass:} The trilinear RPV couplings in Eqs.~\eqref{Eq.lambda_prime} and \eqref{Eq.RPVLLE} contribute to neutrino masses at one-loop level through the lepton-slepton and quark-squark loops~\cite{Hall:1983id, Babu:1989px, Barbier:2004ez}. To ensure that the neutrino masses remain small and satisfy the cosmological bound on the sum $\sum_i m_{\nu_i}\lesssim 0.1$ eV~\cite{Aghanim:2018eyx}, we require some degree of cancellation between the soft trilinear $A$-terms and the $\mu\tan\beta$ term~\cite{Altmannshofer:2020axr}, depending on the other SUSY parameters. \\
    \noindent
    $\bullet$ {\bf Dark Matter:} In RPV scenarios, the lightest supersymmetric particle (LSP) is no longer stable, but decays to SM particles. Therefore, it cannot be the dark matter of the Universe, unless it is sufficiently long-lived, which requires extremely small values of the RPV couplings. In our RPV3 scenario with ${\cal O}(1)$ RPV couplings, the neutralino LSP cannot be the dark matter. However, a gravitino LSP with its naturally Planck-suppressed decays can in principle have a lifetime much longer than the age of the universe, and hence, be the dark matter~\cite{Moreau:2001sr}. \\
    \noindent
$\bullet$ {\bf Hierarchy of RPV Couplings:} For our 
    numerical analysis, we have treated the relevant RPV couplings as free parameters and find the best-fit that explains the flavor anomalies. We find that some of the RPV couplings need to be fairly large $\gtrsim {\cal O}(1)$, while some others need to be hierarchically smaller, and yet others need to be extremely small or  vanishing. One could in principle invoke a flavor symmetry (similar to Ref.~\cite{Trifinopoulos:2018rna} for instance) to explain such hierarchy between couplings; so this need  not be an insurmountable issue, although it would require further work.

\section{Conclusion} \label{sec:con}
The flavor anomalies might already be giving us the first glimpse of natural supersymmetry with light third-generation sfermions and with $R$-parity violating couplings. We have proposed a simple, testable RPV3 scenario that simultaneously explains the $R_{D^{(*)}}$, $R_{K^{(*)}}$ and $(g-2)_\mu$ anomalies with TeV-scale sbottom and tau sneutrino which are easily accessible at the HL-LHC. With experimental updates from LHCb, Belle-II and Fermilab muon $(g-2)$ experiments in the next few years, as well as with better limits on third-generation sfermion masses from the LHC, our knowledge of the anomalies will surely evolve and the allowed RPV3 ranges shown in Figs.~\ref{fig:allowed} and \ref{fig:gm} may have to be modified accordingly. But let us hope SUSY prevails in the end.      

%\end{enumerate}

%\section{Conclusions} \label{sec:con}
%{\bf Conclusions:} 
%We have explicitly shown that a minimal RPV-SUSY scenario with only third-generation sfermions in the low-energy spectrum provides a natural, well-motivated framework to address all the flavor anomalies, while being consistent with current experimental constraints from both low-energy and high-energy sectors. 

%We have also identified striking collider signals that can be used as an independent high-$p_T$ test of the RPV3 interpretation of the flavor anomalies. In particular, the $\bar{t}\mu^+\mu^-$ and four-muon final states proposed here can completely cover the allowed parameter space that explains all the flavor anomalies in the RPV3 scenario.    

%\section*{Acknowledgments}
\acknowledgments
We gratefully acknowledge many useful discussions  with Wolfgang Altmannshofer and Yicong Sui. BD would like to thank Marcela Carena for an important suggestion that led to our Fig.~\ref{fig:scan}. BD also thanks Lawrence Hall for suggesting the $\lambda_{233}$ scenario for $(g-2)_\mu$. 
AS would like to thank Mattia Bruno, Christoph Lehner, Kalman Szabo and Hartmut Wittig for helpful discussions on the SM calculation of muon $g-2$ and related issues.  AS also thanks Yoav Afik for useful discussion on LHC searches relevant for RPV3. The work of BD and FX is supported in part by the U.S. Department  of  Energy  under  Grant  No.   DE-SC0017987 and by the MCSS funds. BD is also supported in part by the Neutrino Theory  Network  Program  and by a Fermilab Intensity Frontier Fellowship. The work of AS was supported in part by the  U.S.  DOE  contract  \#DE-SC0012704.

\appendix
 \section{Low energy constraints} \label{app:constraints}
Despite the many free parameters our RPV3 scenario is remarkably well-constrained by various low-energy flavor observables so much so that
more accurate measurements of $R_{D^{(*)}}$, $R_{K^{(*)}}$ and $(g-2)_\mu$ preserving the central values could have appreciable adverse consequences on our RPV3 explanation of these anomalies. In this section, we summarize all relevant constraints on our RPV3 scenario shown in Fig.~\ref{fig:allowed}. For more details and additional constraints (which are weaker, and therefore not mentioned here), see Ref.~\cite{Altmannshofer:2020axr}.

\subsection{\texorpdfstring{$B \to K^{(*)} \nu \overline{\nu}$}{}}
%%%%%%%%%%%%%%%%%%%%%%%%%%%%%%%%%%%%%%%%%%%%%%%%%%
%%%%%%%%%%%diagram for B to K nu nu
\begin{figure}[tbh!]
		\centering
		\includegraphics[width=0.49\linewidth]{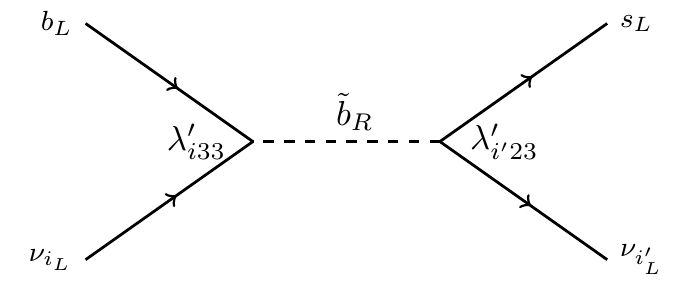}  
		\includegraphics[width=0.49\linewidth]{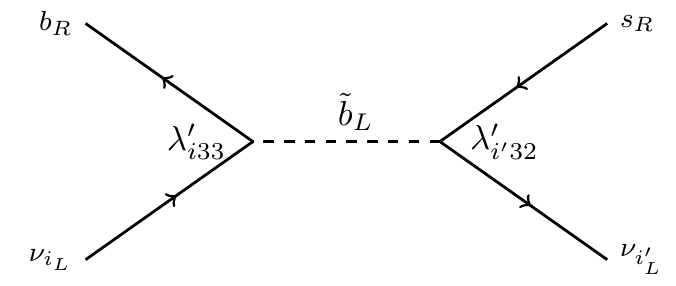}  
	\caption{Contributions to $B\to K^{(*)}\nu \overline{\nu}$ via $\lambda'$ interactions in RPV3.}
	\label{fig:B2Knunu}
\end{figure}
As shown in Fig.~\ref{fig:B2Knunu}, $B \to K^{(*)} \nu \overline{\nu}$ receives a tree-level RPV3 contribution via sbottom exchange. The branching ratio is given by 
\begin{eqnarray} 
 && R_{B\to K^{(*)}\nu \overline{\nu}} \ \equiv \  \frac{{\rm BR}(B \to K^{(*)} \nu \overline{\nu})}{{\rm BR}(B \to K^{(*)} \nu \overline{\nu})_\text{SM}} \nonumber \\
  && \ = \ \frac{1}{3}\left| \delta_{ii'} - \frac{v^2 \pi s_w^2}{2 \alpha_\text{em}} \frac{\lambda^\prime_{i33} }{V_{tb}V_{ts}^*} \left(\frac{\lambda^\prime_{i'23}}{m_{\widetilde b_{R}}^2}+\frac{\lambda^\prime_{i'32}}{m_{\widetilde b_{L}}^2}\right) \frac{1}{X_t}  \right|^2.
%  && ~~ = \frac{1}{3} + \frac{1}{3}\left| 1 + \frac{v^2}{2 m_{\widetilde b_R}^2} \frac{\pi s_w^2}{\alpha_\text{em}} \frac{\lambda^\prime_{333} \lambda^\prime_{323}}{V_{tb}V_{ts}^*} \frac{1}{X_t}  \right|^2  \nonumber \\
%  && ~~ + \frac{1}{3}\left| 1 + \frac{v^2}{2 m_{\widetilde b_R}^2} \frac{\pi s_w^2}{\alpha_\text{em}} \frac{\lambda^\prime_{233} \lambda^\prime_{223}}{V_{tb}V_{ts}^*} \frac{1}{X_t}  \right|^2  \nonumber \\
%  && ~~ + \frac{1}{3}\frac{v^4}{4 m_{\widetilde b_R}^4} \frac{\pi^2 s_w^4}{\alpha_\text{em}^2} \frac{|\lambda^\prime_{233} \lambda^\prime_{323}|^2 + | \lambda^\prime_{333}\lambda^\prime_{223}|^2}{|V_{tb}V_{ts}^*|^2} \frac{1}{X^2_t} \nonumber\\
%  && ~~+ \frac{1}{3}\frac{v^4}{4 m_{\widetilde b_R}^4} \frac{\pi^2 s_w^4}{\alpha_\text{em}^2} \frac{|\lambda^\prime_{233} \lambda^\prime_{123}|^2 +  |\lambda^\prime_{333} \lambda^\prime_{123}|^2}{|V_{tb}V_{ts}^*|^2} \frac{1}{X^2_t} 
\label{eq:RBKvv}
\end{eqnarray}
with the top loop function $X_t = 1.469\pm0.017$ \cite{Brod:2010hi} and $s_w$ being the weak mixing angle. We consider both $\widetilde b_L$ and $\widetilde b_R$ exchanges assuming that $m_{\widetilde{b}_L}=m_{\widetilde{b}_R}$ for numerical purposes. An experimental upper bound for this ratio exists: $R_{B\to K^{(*)}\nu \overline{\nu}} < 5.2$ at 95\% CL~\cite{Buttazzo:2017ixm,Bordone:2018hqs}, which was adopted for our original parameter setting and indicated in Fig.~\ref{fig:allowed} as the solid brown line. However, stronger upper bounds of $R_{B\to K\nu \overline{\nu}} < 3.9$ and $R_{B\to K^{*}\nu \overline{\nu}} < 2.7$ have been quoted by Belle but at 90\% CL~\cite{Belle:2017oht}. In order to make a fair comparison with the other low-energy and collider bounds which are all given at 95\% CL,  we have derived an approximate $95\%$ CL equivalent bound using the Belle data provided in Ref.~\cite{Belle:2017oht}. We get $R_{B\to K^*\nu\overline{\nu}} \lesssim 3.2$, where we have used the theoretical uncertainty from Ref.~\cite{Buras:2014fpa} and have also taken into account the propagation of uncertainty. This $95\%$ CL upper limit is shown in Fig.~\ref{fig:allowed} by the dashed brown line.

\subsection{\texorpdfstring{$B_s-\overline{B}_s$ Mixing}{Bs Bs-bar mixing}}
%%%%%%%%%%%%%%%%%%%%%%%%%%%%%%%%%%%%%%%%%%%%%%%%%%
%%%%%%%%%%%diagram for BB mixing
\begin{figure}[tbh!]
		\centering
		\includegraphics[width=0.49\linewidth]{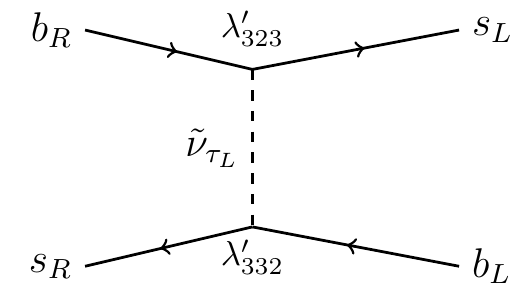}  \\
		\includegraphics[width=0.49\linewidth]{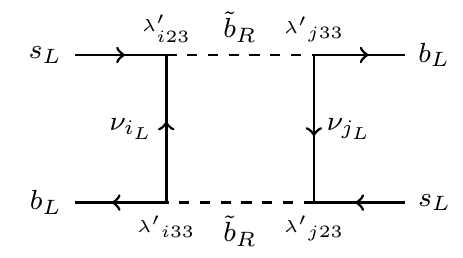}  
		\includegraphics[width=0.49\linewidth]{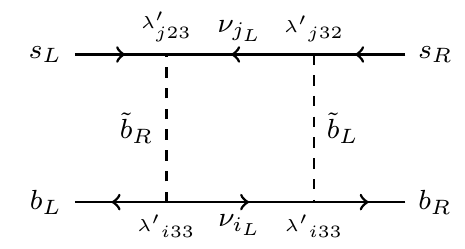}  
	\caption{Relevant contributions to $B_s-\overline{B}_s$ mixing via $\lambda'$ couplings in RPV3.}
	\label{fig:BsBsbar}
\end{figure}
Experimentally, the mass difference $\Delta M_{B_s}$ in neutral $B_s$ meson mixing is measured with excellent precision, $\Delta M_{B_s} = (17.757 \pm 0.021)$\,ps$^{-1}$~\cite{Amhis:2019ckw},  dominated by LHCb and still statistically limited. On the other hand, the SM prediction $\Delta M_{B_s}^\text{SM} = (19.3 \pm 1.7)$\,ps$^{-1}$~\cite{DiLuzio:2019jyq, Altmannshofer:2020axr} has sizable uncertainties stemming mainly from the hadronic matrix elements and the CKM matrix element $V_{cb}$. 

In RPV3, additional contributions can arise at the tree level from sneutrino exchange, or at the one-loop level from box diagrams with sbottoms, sneutrinos, or stops (see Fig.~\ref{fig:BsBsbar}). For the mass difference, we obtain
\begin{align}\label{eq:bbmixing}
    &\Delta M_{B_s}^{\rm RPV} \ = \ \frac{2}{3} m_{B_s}
     f_{B_s}^2\left|P^{VLL}_1 \frac{\lambda'_{i23}\lambda'_{j33}\lambda'_{j23}\lambda'_{i33}}{128\pi^2 m_{\widetilde b_R}^2} \right. \nonumber\\ &\quad \left. +P^{LR}_1 \frac{\lambda'_{i23}\lambda'_{j33}\lambda'_{i32}\lambda'_{j33}}{128\pi^2 m_{\widetilde b_R}m_{\widetilde b_L}} + P^{LR}_2 \frac{\lambda'_{332}\lambda'_{323}}{2 m_{\widetilde \nu}^2}
    \right| \, ,
\end{align}
where 
\begin{equation}
    P^{VLL}_1 \ = \ 0.80\, , \, 
    P^{LR}_1 \ = \ -2.52\, \, {\rm and}\, \, P^{LR}_2 \ = \ 3.08~,
\end{equation}
are the updated hadronic $P$ factors from  Ref.~\cite{Buras:2001ra} with the latest lattice input from Ref.~\cite{Bazavov:2016nty} (see also Refs.~\cite{Boyle:2018knm, FlavourLatticeAveragingGroup:2019iem}), $ f_{B_s} = ( 274 \pm 8 )$\,MeV is the $B_s$ decay constant, and $i,j$ are neutrino-flavor indices in the box graphs. Combining our SM prediction with the experimental result we obtain the following bound at 95\% C.L. on 
\begin{equation}
 0.78 \ < \ \left| \frac{\Delta M_{B_s}}{\Delta M_{B_s}^\text{SM}} \right| \ < \ 1.12 ~,
\end{equation}
which constrains the RPV3 contribution in Eq.~\eqref{eq:bbmixing}. This bound is indicated as the magenta-shaded region in  Fig.~\ref{fig:allowed}.

\subsection{\texorpdfstring{$D^0 \to \mu^+\mu^-$}{D0 -> mu+mu-}} \label{Dtomumu}
%%%%%%%%%%%%%%%%%%%%%%%%%%%%%%%%%%%%%%%%%%%%%%%%%%
%%%%%%%%%%%diagram for D to mu mu 
\begin{figure}[tbh!]
		\centering
		\includegraphics[width=0.25\textwidth]{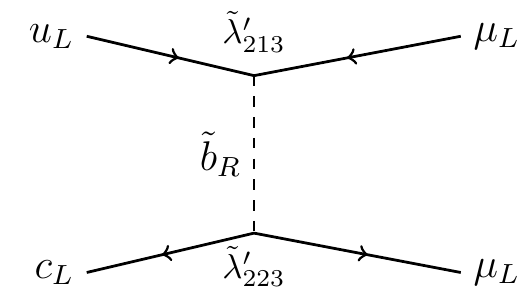}
	\caption{Contribution to $D^0 \to \mu^+\mu^-$ from $\lambda'$ in RPV3.}
	\label{fig:Dmumu}
\end{figure}
As shown in Fig.~\ref{fig:Dmumu}, there is a tree-level contribution from sbottom exchange to this rare $D^0$ decay width which can be expressed as
\begin{align}
    \Gamma(D^0 \to \mu^+ \mu^-) & \ = \  \frac{1}{128\pi}\left|\frac{\lambda'_{2j3}\lambda'_{2j'3} V_{uj'} V_{cj}}{m_{\widetilde b_R}^2}\right|^2 f_D^2 \nonumber\\
   &\times m_D m_{\mu}^2\sqrt{1-4m_{\mu}^2/m_D^2} \, , 
\end{align}
where $f_D=(212\pm 1)$ MeV is the $D^0$ decay constant. 
Using the experimental upper bound on this branching ratio \cite{Aaij:2013cza} of $7.6\times 10^{-9}$ at 95\% CL, we calculate the corresponding bound on the RPV3 parameter space, as shown by the purple-shaded region in Fig.~\ref{fig:allowed}. 

\subsection{\texorpdfstring{$Z\to \ell \overline{\ell}'$}{Z -> l l'-bar}}
%%%%%%%%%%%%%%%%%%%%%%%%%%%%%%%%%%%%%%%%%%%%%%%%%%
%%%%%%%%%%%diagram for Z to ll
\begin{figure}[tbh!]
		\centering
		\includegraphics[width=0.26\textwidth]{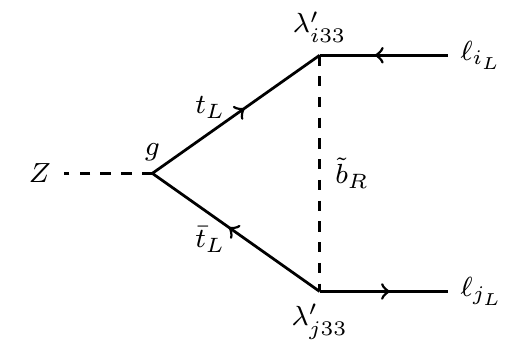}
	\caption{Contribution to $Z\to \ell \overline{\ell}'$ from $\lambda'$ in RPV3.}
	\label{fig:Zll}
\end{figure}
This process gets modified by top-sbottom loops, as shown in Fig.~\ref{fig:Zll}. 
%More specifically, $Z$ will decay to an off-shell $t\bar t$ pair both of which through $LQD$ give a lepton and a sbottom. The sbottom line closes to form a loop, while the product of the process becomes a $l\bar l'$ final state. 
A change in the $Z$ decay from the SM prediction will affect the ratios of the vector and axial-vector couplings of the $Z$ boson with different lepton flavors. Experimental measurements on these couplings are~\cite{ParticleDataGroup:2020ssz} 
\begin{align}
    \left(\frac{g_V^\tau}{g_V^e}\right)_{\rm exp} & \ = \  0.9588 \pm 0.02997 \, , \label{eq:zll1}\\
    \left(\frac{g_A^\tau}{g_A^e}\right)_{\rm exp} & \ = \  1.0019 \pm 0.00145 \, . \label{eq:zll2}
\end{align}
The contributions to these ratios from RPV3 are given by
\begin{align}
    \left(\frac{g_V^\tau}{g_V^e}\right)_{\rm SM+RPV} & \ = \ 1 - \frac{2\,\delta g_{\ell_3 \ell_3}}{1-4\,s_w^2} \, , \nonumber\\
    \left(\frac{g_A^\tau}{g_A^e}\right)_{\rm SM+RPV} & \ = \ 1 - 2\,\delta g_{\ell_3 \ell_3} \, ,\nonumber
\end{align}
where 
\begin{align}
    \delta g_{\ell_i \ell_j} & \ \simeq \ \frac{3 y_t^2}{32\sqrt{2}G_F \pi^2} \frac{\lambda'_{i33} \lambda'_{j33}}{m_{\widetilde b_R}^2}\left[\log\left(\frac{m_{\widetilde b_R}}{m_Z}\right)-0.612\right] \, .
\end{align}
Taking $i$, $j$ both equal to 3 and using Eqs.~\eqref{eq:zll1} and \eqref{eq:zll2}, we derive a bound on the RPV3 parameter space, as shown by the violet-shaded region in Fig.~\ref{fig:allowed}.

\subsection{\texorpdfstring{$b \to s \gamma$}{b -> s gamma}} \label{sec:bsgamma}
%%%%%%%%%%%%%%%%%%%%%%%%%%%%%%%%%%%%%%%%%%%%%%%%%%
%%%%%%%%%%%diagram for b to s gamma
\begin{figure}[tbh!]
		\centering
		\includegraphics[width=.23\textwidth]{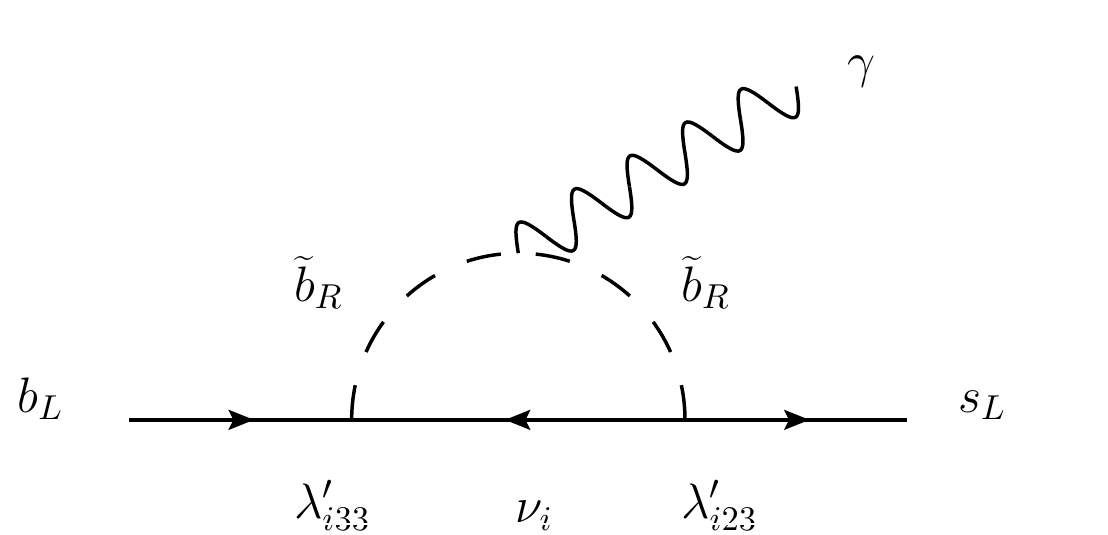}
		\includegraphics[width=.23\textwidth]{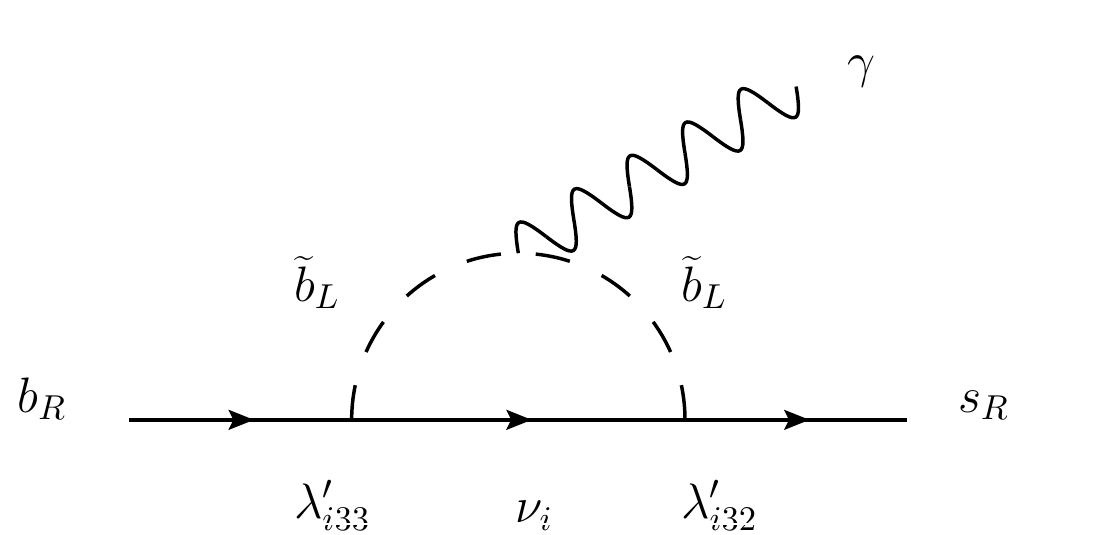}
	\caption{Contribution to $b \to s \gamma$ from $\lambda'$ couplings in RPV3.}
	\label{fig:bsgamma}
\end{figure}
The branching ratio of $b\to s\gamma$ has been measured~\cite{Amhis:2019ckw} as:
\begin{equation}
    {\rm BR}(b\to s\gamma)_{\rm exp} \ =  \ (3.43 \pm 0.21 \pm 0.07) \times 10^{-4}\, ,
\end{equation}
which is consistent with SM~\cite{Misiak:2015xwa}: 
\begin{align}
  {\rm BR}(b\to s\gamma)_{\rm SM} \ = \ (3.36 \pm 0.23)\times 10^{-4}  \, .
\end{align}
In RPV3, there are one-loop contributions involving both left- and right-handed sbottoms (see Fig.~\ref{fig:bsgamma}). Comparing this to the difference between the experimental and SM results, we obtain the following bound at 95\% CL: 
\begin{align}
    |\lambda^\prime_{223} \lambda^\prime_{233}|& \ \lesssim \ 0.025 \left( \frac{100 \rm{\ GeV}}{m_{\widetilde{b}_R}} \right)^{-2} \\
    |\lambda^\prime_{232} \lambda^\prime_{233}|& \ \lesssim \ 0.01 \left( \frac{100 \rm{\ GeV}}{m_{\widetilde{b}_L}} \right)^{-2} %\\
   % \lambda'_{323}\lambda'_{333}|& \ \lesssim \ 0.025 \left| 2\left(\frac{100\ {\rm GeV}}{m_{\widetilde \nu_\tau}}\right)^2 -\left(\frac{100\ {\rm GeV}}{m_{\widetilde b_{R}}}\right)^2 \right|^{-1} , \label{eq:bsg1} \\
    %|\lambda'_{332}\lambda'_{333}|& \ \lesssim \ 0.01 \left| \left(\frac{100 \ {\rm GeV}}{m_{\widetilde \tau_{L}}}\right)^2 -\left(\frac{100 \ {\rm GeV}}{m_{\widetilde b_{L}}}\right)^2 \right|^{-1}. \label{eq:bsg2}
\end{align}
This is shown by the grey-shaded region in Fig.~\ref{fig:allowed}.  

\bibliographystyle{apsrev4-1}
\bibliography{references}

\end{document}